\documentclass[lettersize,journal]{IEEEtran}
\IEEEoverridecommandlockouts % to insert the /thanks{} section in the bottom left corner for conference papers
\usepackage{caption}
\usepackage{subcaption}
\usepackage{enumerate}
\usepackage{graphicx}
\usepackage{enumitem}
\usepackage{psfrag}
\usepackage{acronym}
\usepackage{tikz}
\usepackage{pgfplots}
\usepackage{xparse}
\usepackage{cite}
\usepackage{amsmath,amssymb,amsfonts}
\usepackage{graphicx}
\usepackage{xcolor}
\usepackage{colortbl} % for colored rows
\usepackage{hhline} % for double horizontal lines
\usepackage{soul}
\usepackage{array}
\usepackage{verbatim}
\usepackage{bm}
\usepackage{multirow}
\usepackage{bbm}
\usepackage{tabularx}
\usepackage{amsthm} 
\usepackage{comment}

\usepackage[ruled,vlined]{algorithm2e}
\SetAlFnt{\small}                 % font più compatto
\SetAlCapFnt{\small}
\SetAlCapNameFnt{\small}
\SetAlgoNlRelativeSize{-1}        % numerazione meno invadente
\SetAlCapHSkip{0pt}

\SetKwInput{KwIn}{Input}
\SetKwInput{KwOut}{Output}
\SetAlgoLined
\LinesNumbered
 

\usepackage{cleveref}
\crefname{algocf}{Algorithm}{Algorithms}
\Crefname{algocf}{Algorithm}{Algorithms}
\DontPrintSemicolon      % per togliere i ; di default a fine riga
\SetKwRepeat{Do}{do}{while}   % per usare i loop tipo do...while

\begin{document}

\title{OAM-Enabled Holographic MIMO Communications with Stacked Intelligent Metasurfaces}
\author{\IEEEauthorblockN{G. Torcolacci,~\IEEEmembership{ Member,~IEEE}, and D. Dardari,~\IEEEmembership{Fellow,~IEEE}} }

\markboth{OAM-Enabled Holographic MIMO Communications with Stacked Intelligent Metasurfaces}
{Torcolacci \MakeLowercase{\textit{et al.}}}

\maketitle

\begin{abstract}
This study investigates orbital angular momentum (OAM)-based holographic multiple-input multiple-output (HMIMO) links enabled by stacked intelligent metasurfaces (SIM) in the radiative near-field. By using multilayer programmable metasurfaces at both link ends, SIMs enable analog electromagnetic domain wave processing for low-complexity and energy-efficient flexible wavefront synthesis. We analyzed OAM mode generation and reception with SIM-based transceivers and quantified their ability to synthesize near-orthogonal modes in practical discrete HMIMO architectures. We further developed a correlation-driven optimization algorithm that maximizes reconstruction accuracy of OAM beams. Numerical evaluations revealed a fundamental decoupling between the required antenna aperture, which limits the supported mode orders, and the SIM layer depth, which governs crosstalk suppression. The results confirm that properly dimensioned SIM architectures provide robust near-field spatial multiplexing, nearly balanced per-mode capacities, and graceful degradation across link distances without requiring continuous phase re-optimization, thereby supporting scalable and low-overhead HMIMO communications.
\end{abstract}


% ACRONYMS
%
%
% by Giulia Torcolacci

%\begin{acronym}

% usage: \ac{SW}, \acp{SW} for plurals \acf{SW} Use the full name of the acronym.
%\acs{SW}Use the acronym, even before the first corresponding \ac command
%\acl{acronym}Expand the acronym without using the acronym itself.
%\scriptsize

% MATH ACRONYMS
\acrodef{$P_{EM}$}{probability of emulation, or false alarm}
\acrodef{$P_{FA}$}{probability of false alarm}
\acrodef{$P_{MD}$}{probability of missed detection}
\acrodef{$P_{D}$}{probability of detection}

% TEXT ACRONYMS
\acrodef{2D}{bi-dimensional}
\acrodef{3D}{three-dimensional}
\acrodef{5G}{5th generation}
\acrodef{6G}{6th Generation}
\acrodef{ACF}{autocorrelation function}
\acrodef{ACG}{automatic	gain control}
\acrodef{ACI}{adjacent channel interference}
\acrodef{ACK}{acknowledge}
\acrodef{AcR}{autocorrelation receiver}
\acrodef{Adam}{adaptive moment estimation}
\acrodef{ADC}{analog-to-digital converter}
\acrodef{AF}{amplify \& forward}
\acrodef{AFL}{anchor-free localization}
\acrodef{AGNSS}{assisted-GNSS}
\acrodef{AGPS}{assisted GPS}
% \acrodef{AI}{automatic identification}
\acrodef{AI}{artificial intelligence}
\acrodef{AIC}{Akaike information criterion}
\acrodef{AO}{alternating optimization}
\acrodef{AOA}{angle-of-arrival}
\acrodef{AOD}{angle-of-departure}
\acrodef{AOT}{approximate optimum threshold}
\acrodef{AP}{access point}
\acrodef{API}{application programming interface}
\acrodef{ASK}{amplitude shift keying}
\acrodef{ASNR}{accumulated signal-to-noise ratio}
\acrodef{AUB}{asymptotic union bound}
\acrodef{AWGN}{additive white Gaussian noise}
\acrodef{BAN}{body area network}
\acrodef{BAV}{balanced antipodal Vivaldi}
\acrodef{BCH}{Bose Chaudhuri Hocquenghem}
\acrodef{BD-RIS}{beyond-diagonal reconfigurable intelligent surfaces}
\acrodef{BEP}{bit error probability}
\acrodef{BER}{bit error rate}
\acrodef{BF}{brute force}
\acrodef{BFC}{block fading channel}
\acrodef{BIC}{Bayesian information criterion}
\acrodef{BLUE}{best linear unbiased estimator}
\acrodef{BPAM}{binary pulse amplitude modulation}
\acrodef{BPF}{bandpass filter}
\acrodef{BPPM}{binary pulse position modulation}
\acrodef{bps}{bits per second}
\acrodef{BPSK}{binary phase shift keying}
\acrodef{BPZF}{band-pass zonal filter}
\acrodef{BS}{base station}
\acrodef{BSC}{binary symmetric channel}
\acrodef{BTB}{Bellini-Tartara bound}
\acrodef{c.c.d.f.}{complementary cumulative distribution function}
\acrodef{c.d.f.}{cumulative distribution function}
\acrodef{CAD}{computer-aided design}
\acrodef{CAIC}{consistent Akaike information criterion}
\acrodef{CAP}{continuous aperture phased}
\acrodef{CCF}{cross correlation function}
\acrodef{CCI}{co-channel interference}
\acrodef{CD}{cooperative diversity}
\acrodef{CDMA}{code division multiple access}
\acrodef{CEOT}{channel ensemble optimum threshold}
\acrodef{CEP}{codeword error probability}
\acrodef{CFAR}{constant	 false alarm rate}
\acrodef{ch.f.}{characteristic function}
\acrodef{CH}{cluster head}
\acrodef{CIR}{channel impulse response}
\acrodef{CL}{centroid localization}
\acrodef{CM}{channel model}
\acrodef{CNR}{clutter-to-noise ratio}
\acrodef{CP}{ciclic prefix}
\acrodef{CPSF}{circular prolate spheroidal function}
\acrodef{CPR}{channel pulse response}
\acrodef{CR}{channel response}
\acrodef{CRB}{Cram\'{e}r-Rao bound}
\acrodef{CRC}{cyclic redundancy check}
\acrodef{CRLB}{Cram\'{e}r-Rao lower bound}
\acrodef{CS}{clock skew}
\acrodef{CSCG}{circularly symmetric complex Gaussian}
\acrodef{CSI}{channel state information}
\acrodef{CSMA}{carrier sense multiple access}
\acrodef{CSS}{chirp spread spectrum}
\acrodef{CTS}{clear-to-send}
\acrodef{CW}{continuous wave}
\acrodef{DAA}{detect and avoid}
\acrodef{DAB}{digital audio broadcasting}
\acrodef{DAC}{digital-to-analog converter}
\acrodef{DMA}{dynamic metasurface antenna}
\acrodef{DBB}{digital base band}
\acrodef{DBPSK}{differential binary phase shift keying}
\acrodef{DCM}{dual-carrier modulation}
\acrodef{DDP}{detected direct path}
\acrodef{DF}{detect \& forward}
\acrodef{DFMS}{monopole dual feed stripline antenna}
\acrodef{DFT}{discrete Fourier transform}
\acrodef{DGPS}{differential GPS}
\acrodef{DLL}{delay-locked loop}
\acrodef{DNN}{deep neural network}
\acrodef{DOA}{direction of arrival}
\acrodef{DoD}{Department of Defense}
\acrodef{DOF}{degrees of freedom}
\acrodef{DP}{direct path}
\acrodef{DR}{detection rate}
\acrodef{DRT}{distance ratio test}
\acrodef{DS-SS}{direct-sequence spread-spectrum}
\acrodef{DS}{delay spread}
\acrodef{DTR}{differential transmitted-reference}
\acrodef{DTT}{Diffraction Tomography Theory}
\acrodef{DVB-H}{digital video broadcasting\,--\,handheld}
\acrodef{DVB-T}{digital video broadcasting\,--\,terrestrial}
\acrodef{e.m.}{electromagnetic}
\acrodef{ECC}{European Community Commission}
\acrodef{ED}{energy detector}
\acrodef{EDR}{energy detector receiver}
\acrodef{EFIM}{equivalent Fisher information matrix}
\acrodef{EIRP}{effective radiated isotropic power}
\acrodef{ESP}{electromagnetic signal processing}
\acrodef{EKF}{extended Kalman filter}
\acrodef{KKT}{Karush–Kuhn–Tucker}
\acrodef{ELP}{equivalent low-pass}
%\acrodef{EM}{expectation-maximization}
\acrodef{EM}{electromagnetic}
\acrodef{EMCB}{extended Miller Chang bound}
\acrodef{EME}{minimum eigenvalue ratio detector}
\acrodef{EMI}{electromagnetic interference}
\acrodef{ENP}{estimated noise power}
\acrodef{ESA}{European Space Agency}
\acrodef{EU}{European Union}
\acrodef{EVD}{eigenvalue decomposition}
\acrodef{FAR}{false alarm rate}
\acrodef{FCC}{Federal Communications Commission}
\acrodef{FDMA}{frequency division multiple access}
\acrodef{FDMA}{frequency division multiple access}
\acrodef{FEC}{forward error correction}
\acrodef{FEC}{forward error correction}
\acrodef{FFD}{full function device}
\acrodef{FFR}{full function reader}
\acrodef{FF}{far-field}
\acrodef{FFT}{fast Fourier transform}
\acrodef{FG}{factor graph}
\acrodef{FH-SS}{frequency-hopping spread-spectrum}
\acrodef{FH}{frequency-hopping}
\acrodef{FIM}{Fisher information matrix}
\acrodef{FLL}{Frequency-locked loop}
\acrodef{FPGA}{field programmable gate array}
\acrodef{FS}{frame synchronization}
\acrodef{FT}{Fourier Transform}
\acrodef{GA}{Gaussian approximation}
\acrodef{GD}{gradient descent}
\acrodef{GDOP}{geometric dilution of precision}
\acrodef{GLR}{generalized likelihood ratio}
\acrodef{GLRT}{generalized likelihood ratio test}
\acrodef{GML}{generalized maximum likelihood}
\acrodef{GPRS}{general packet radio service}
\acrodef{GPS}{global positioning system}
\acrodef{HAP}{high altitude platform}
\acrodef{HCRB}{hybrid Cram\'{e}r-Rao bound}
\acrodef{HDSA}{high-definition situation-aware}
\acrodef{Hi-RADIAL}{High-accuracy RAdio Detection, Identification, And Localization}
\acrodef{HMIMO}{holographic multiple-input multiple-output}
\acrodef{HMM}{hidden Markov model}
\acrodef{HPA}{high-power amplifier}
\acrodef{HPBW}{half power beam width}
\acrodef{HW}{hardware}
\acrodef{i.i.d.}{independent, identically distributed}
\acrodef{ICT}{information and communication technologies}
\acrodef{IE}{informative element}
\acrodef{IEEE}{Institute of Electrical and Electronics Engineers}
\acrodef{IF}{intermediate frequency}
\acrodef{IFFT}{inverse fast Fourier transform}
\acrodef{IMF}{ideal matched filter}
\acrodef{IMU}{inertial measurement unit}
\acrodef{INR}{interference-to-noise ratio}
\acrodef{INS}{inertial navigation system}
\acrodef{IoT}{Internet of things}
\acrodef{IIoT}{industrial Internet of things}
\acrodef{INS}{inertial navigation system}
\acrodef{IR-UWB}{impulse radio UWB}
\acrodef{IR}{impulse radio}
\acrodef{IRI}{inter-reader interference}
\acrodef{IRS}{intelligent reflecting surface} 
\acrodef{ISAC}{integrated sensing and communication}
\acrodef{ISI}{inter-symbol interference} 
\acrodef{isi}{intra-symbol interference} 
\acrodef{ISM}{industrial, scientific and medical}
\acrodef{ISNR}{interference-plus-signal-to-noise-ratio}
\acrodef{ISP}{inverse scattering problem}
\acrodef{IT}{interference temperature}
\acrodef{ITC}{information theoretic criteria}
\acrodef{JBSF}{jump back and search forward}
\acrodef{JF}{just forward}
\acrodef{KF}{Kalman filter}
\acrodef{KKT}{Karush–Kuhn–Tucker}
\acrodef{LDC}{low duty cycle}
\acrodef{LDPC}{low density parity check}
\acrodef{LEO}{localization error outage}
\acrodef{LG}{Laguerre-Gaussian}
\acrodef{LIS}{large intelligent surface}
\acrodef{LLR}{log-likelihood ratio}
\acrodef{LLRT}{log-likelihood ratio test}
\acrodef{LR}{long-range}
\acrodef{LRT}{likelihood ratio test}
\acrodef{LNA}{low-noise amplifier}
\acrodef{LOS}{line-of-sight}
\acrodef{LRT}{likelihood ratio test}
\acrodef{LS}{least square}
\acrodef{LS}{least squares}
\acrodef{M-PSK}{$M$-ary phase shift keying}
\acrodef{M-QAM}{$M$-ary quadrature amplitude modulation}
\acrodef{m.g.f.}{moment generating function}
\acrodef{MAC}{medium access control}
\acrodef{MAE}{mean absolute error}
\acrodef{MAI}{multiple access interference}
\acrodef{MAN}{metropolitan area network}
\acrodef{MAP}{maximum a posteriori}
\acrodef{MB-OFDM}{multi-band OFDM}
\acrodef{MB-UWB}{multi-band UWB}
\acrodef{MB}{multi-band}
\acrodef{MC}{multi-carrier}
\acrodef{MCB}{Miller Chang bound}
\acrodef{MCRB}{modified Cram\'{e}r-Rao bound}
\acrodef{MDD}{minimum distance distribution}
\acrodef{MDL}{minimum description length}
\acrodef{MF}{matched filter}
\acrodef{MGF}{moment generating function}
\acrodef{MI}{mutual information}
\acrodef{MIMO}{multiple-input multiple-output}
\acrodef{MISO}{multiple-input single-output}
\acrodef{ML}{maximum likelihood}
\acrodef{MM}{min-max}
\acrodef{MME}{maximum-minimum eigenvalue ratio detector}
\acrodef{mmWave}{millimeter wave}
\acrodef{MMSE}{minimum mean-square error}
\acrodef{MPC}{multipath component}
\acrodef{MRC}{maximal ratio combiner}
\acrodef{MS}{mobile station}
\acrodef{MSB}{most significant bit}
\acrodef{MSE}{mean squared error}
\acrodef{NMSE}{normalized mean squared error}
\acrodef{MSK}{minimum shift keying}
\acrodef{MU}{multiuser}
\acrodef{MUI}{multi-user interference}
\acrodef{MUR}{multistatic radar}
\acrodef{MVU}{minimum variance unbiased}
\acrodef{MZZB}{modified Ziv-Zakai bound}
\acrodef{NB}{narrowband}
\acrodef{NBI}{narrowband interference}
\acrodef{NEO}{navigation error outage}
\acrodef{NFER}{near-Þeld electromagnetic ranging}
\acrodef{NF}{near-field}
\acrodef{NFF}{near-field focused}
\acrodef{NL}{nonlinear}
\acrodef{NLOS}{non-line-of-sight}
\acrodef{NP}{Neyman-Pearson}
\acrodef{NTIA}{National Telecommunications and Information Administration}
\acrodef{NTP}{network time protocol}
\acrodef{OAM}{orbital angular momentum} 
\acrodef{OC}{optimum combining}
\acrodef{OFDM}{orthogonal frequency division multiplexing}
\acrodef{OOK}{on-off keying}
\acrodef{OP}{outage probability}
\acrodef{OT}{optimum threshold}
\acrodef{P-Max}{$P$-Max}  %suggestion, use with \acl{P-Max}
\acrodef{p.d.f.}{probability density function}
\acrodef{p.m.f.}{probability mass function}
\acrodef{PA}{power amplifier}
\acrodef{PAM}{pulse amplitude modulation}
\acrodef{PAN}{personal area network}
\acrodef{PAR}{peak-to-average ratio}
\acrodef{P-CRLB}{Posterior Cramer-Rao Lower Bound}
\acrodef{PCA}{principal component analysis}
\acrodef{PD}{probability of detection}
\acrodef{PDP}{power delay profile}
\acrodef{PE}{probability of emulation}
\acrodef{PEB}{position error bound}
\acrodef{PEC}{perfect electric conductor}
\acrodef{PEP}{packet error probability}
\acrodef{PF}{particle filter}
\acrodef{PFA}{probability of false alarm}
\acrodef{PHY}{physical layer}
\acrodef{PL}{path-loss}
\acrodef{PLL}{phase-locked loop}
\acrodef{PMD}{probability of missed detection}
\acrodef{PN}{pseudo-noise}
\acrodef{PSF}{point spread function}
\acrodef{ppm}{part-per-million}
\acrodef{PPM}{pulse position modulation}
\acrodef{PR}{pseudo-random}
\acrodef{PRake}{partial rake}
\acrodef{PRF}{pulse repetition frequency}
\acrodef{PRP}{pulse repetition period}
\acrodef{PSD}{power spectral density}
\acrodef{PSEP}{pairwise synchronization error probability}
\acrodef{PSNR}{peak signal to noise ratio}
\acrodef{PSK}{phase shift keying}
\acrodef{PSVD}{product singular value decomposition}
\acrodef{PSWF}{prolate spheroidal wave function}
\acrodef{PU}{primary user}
\acrodef{QAM}{quadrature amplitude modulation}
\acrodef{QoS}{quality of service}
\acrodef{QPSK}{quadrature phase shift keying}
%\acrodef{r.v.}{random variable}
\acrodef{R.V.}{random variable}
\acrodef{RADAR}{radar}
\acrodef{RCS}{radar cross section}
\acrodef{RDL}{"random data limit"}
\acrodef{REM}{radio environment map}
\acrodef{REO}{ranging error outage}
\acrodef{RF}{radio-frequency}
\acrodef{RFID}{radio-frequency identification}
\acrodef{RFR}{reduced function reader}
\acrodef{RFT}{reduced function tag}
\acrodef{RII}{ranging information intensity}
\acrodef{RIS}{reconfigurable intelligent surface}
\acrodef{rms}{root mean square}
\acrodef{RMSE}{root-mean-square error}
\acrodef{ROC}{receiver operating characteristic}
\acrodef{ROI}{region of interest}
\acrodef{RRC}{root raised cosine}
\acrodef{RSN}{radar sensor network}
\acrodef{RSS}{received signal strength}
\acrodef{RSSI}{received signal strength indicator}
\acrodef{RTLS}{real time locating systems}
\acrodef{RTT}{round-trip time}
\acrodef{S-V}{Saleh-Valenzuela}
\acrodef{SA}{simulated annealing}
\acrodef{SaG}{stop-and-go}
\acrodef{SAR}{synthetic aperture radar}
\acrodef{SBS}{serial backward search}
\acrodef{SBSMC}{serial backward search for multiple clusters}
\acrodef{SCM}{supply chain management}
\acrodef{SCR}{signal-to-clutter ratio}
\acrodef{SEP}{symbol error probability}
\acrodef{SIS}{small intelligent surface}
\acrodef{SFD}{start frame delimiter}
\acrodef{SIM}{stacked intelligent metasurface}
\acrodef{SIMO}{single-input multiple-output}
\acrodef{SINR}{signal-to-interference plus noise ratio}
\acrodef{SIR}{signal-to-interference ratio}
\acrodef{SISO}{single-input single-output}
\acrodef{SNR}{signal-to-noise ratio}
\acrodef{SoC}{system on chip}
\acrodef{SoO}{signal of opportunity}
\acrodef{SoP}{system on package}
\acrodef{SOT}{sub-optimum threshold}
\acrodef{SPAWN}{sum-product algorithm over a wireless network}
\acrodef{SPEB}{squared position error bound}
\acrodef{SPMF}{single-path matched filter}
\acrodef{SPP}{spiral phase plate}
\acrodef{SQNR}{signal-to-quantization-noise ratio}
\acrodef{SR}{short-range}
\acrodef{SRE}{smart radio environment}
\acrodef{SS}{spread spectrum}
\acrodef{ST}{simple thresholding}
\acrodef{SU}{secondary user}
\acrodef{SVD}{singular value decomposition}
\acrodef{SW}{software}
\acrodef{SW}{sync word}
\acrodef{TDE}{time delay estimation}
\acrodef{TDL}{tapped delay line}
\acrodef{TDMA}{time division multiple access}
\acrodef{TDOA}{time difference-of-arrival}
\acrodef{TH}{time-hopping}
\acrodef{THz}{terahertz}
\acrodef{TNR}{threshold-to-noise ratio}
\acrodef{TOA}{Time-of-arrival}
\acrodef{TOF}{time-of-flight}
\acrodef{TPC}{transmit power control}
\acrodef{TR}{transmitted-reference}
\acrodef{TS}{tabu search}
\acrodef{TSVD}{truncated singular value decomposition}
\acrodef{TV}{total variation denoising}
\acrodef{UAV}{unmanned aerial vehicle}
\acrodef{UB}{union bound}
\acrodef{UCA}{uniform circular array}
\acrodef{UDP}{undetected direct path}
\acrodef{UE}{User Equipment}
\acrodef{UHF}{ultra-high frequency}
\acrodef{ULA}{uniform linear array}
\acrodef{ULP}{user location protocol}
\acrodef{UMP}{uniformly most powerful}
\acrodef{UMPI}{uniformly most powerful invariant}
\acrodef{URA}{uniform rectangular array}
\acrodef{UT}{user terminal}
\acrodef{UTC}{coordinated universal time}
\acrodef{UTM}{universal transverse Mercator}
\acrodef{UTRA}{UMTS terrestrial radio access}
\acrodef{UAV}{unmanned aerial vehicle}
\acrodef{UUV}{unmanned underwater vehicle}
\acrodef{UWB}{ultrawide-band}
\acrodef{UWBcap}[UWB]{Ultrawide band}
\acrodef{VFIL}{virtual force iterative localization}
\acrodef{VGA}{variable-gain amplifier}
\acrodef{VNA}{vector network analyzer}
\acrodef{WAF}{wall attenuation factor}
\acrodef{WB}{wideband}
\acrodef{WBI}{wideband interference}
\acrodef{WCL}{weighted centroid localization}
\acrodef{WED}{wall extra delay}
\acrodef{WiMAX} {worldwide interoperability for microwave access}
\acrodef{WLAN}{wireless local area network}
\acrodef{WLS}{weighted least squares}
\acrodef{WMAN}{wireless metropolitan area network}
\acrodef{WPAN}{wireless personal area networks}
\acrodef{WRAPI}{wireless research application programming interface}
\acrodef{WSN}{wireless sensor network}
\acrodef{WSR}{wireless sensor radar}
\acrodef{WSS}{wide-sense stationary}
\acrodef{WWB}{Weiss-Weinstein bound}
\acrodef{WWLB}{Weiss-Weinstein lower bound}
\acrodef{ZZB}{Ziv-Zakai bound}
\acrodef{ZZLB}{Ziv-Zakai lower bound}
\acrodef{XL-MIMO}{extremely large-scale multiple-input multiple-output}
\acrodef{TX}{transmitting}
\acrodef{RX}{receiving}
\acrodef{UPA}{uniform planar array}
%\acrodef{AoA}{angle-of-arrival}
%\acrodef{AoD}{angle-of-departure}
%\acrodef{CCDF}{complementary cumulative distribution function}
%\acrodef{CDF}{cumulative distribution function}
%\acrodef{CR}{cognitive radio}
%\acrodef{DF}{decode and forward}
%\acrodef{DS}{direct sequence}
%\acrodef{FFT}{full function tag}
%\acrodef{i.s.i.}{intra-symbol interference}
%\acrodef{LEO}{low earth orbit}
%\acrodef{MAP}{maximum a posteriori probability}
%\acrodef{MUR}{multistatic RADAR}
%\acrodef{PDF}{probability density function}
%\acrodef{RADAR}{RADAR}
%\acrodef{RMS}{root mean square}
%\acrodef{RV}{random variable}
%\acrodef{SoO}{source of opportunity}
%\acrodef{TDoA}{time difference-of-arrival}
%\acrodef{ToA}{time-of-arrival}
%\acrodef{ToF}{time-of-flight}}

%\end{acronym}

% Macros definitions
%
% Torcolacci Giulia
% 15 June. 2023
%----------------------------------------------------------------

% Math Operators 
\newcommand{\rect}[1] {\text{rect} \left ({#1} \right )}
\newcommand{\sinc}[1] {\text{sinc} \left ({#1} \right )}
\newcommand{\argmax}[1]{\underset{{#1}}{\operatorname{argmax}}}
\newcommand{\argmin}[1]{\underset{{#1}}{\operatorname{argmin}}}
\newcommand{\E}[1] {\mathbb{E}\left\{#1\right\}}
\newcommand{\Var}[1] {\mathbb{V}\left\{#1\right\}}
\newcommand{\Real}[1]{\Re^{#1}}
\newcommand{\floor}[1] {f \left ({#1} \right )}
\def\erfc{{\text{erfc}}}
\def\erf{{\text{erf}}}
\def\inverfc{{\text{inverfc}}}
\newcommand{\rank}{{\rm rank}}
\newcommand{\diag}{\mathrm{diag}}
\newcommand{\degree}{\ensuremath{^\circ}}
\newcommand{\ra}{\rightarrow}
\newcommand{\rf}{\leftarrow}
\newcommand{\cn}{{\mathcal{CN}}} %Complex Gaussian RV
\newcommand{\tr}{\operatorname{tr}}
\newcommand{\Variance}[1]{\operatorname{Var} \left({#1} \right)}

% Acronyms
\newcommand{\SNR}{\text{SNR}}
\newcommand{\TNR}{\mathsf{TNR}}
\newcommand{\sigmaN} {\sigma_{\text{N}}}
\newcommand{\MSE} {\text{MSE}}

% BOLD LETTERS
\newcommand{\bolda}{{\bf a}}
\newcommand{\boldb}{{\bf b}}
\newcommand{\boldbeta}{{\boldsymbol{\beta}}}
\newcommand{\boldc}{{\bf c}}
\newcommand{\boldd}{{\bf d}}
\newcommand{\bolde}{{\bf e}}
\newcommand{\boldf}{{\bf f}}
\newcommand{\boldg}{{\bf g}}
\newcommand{\boldh}{{\bf h}}
\newcommand{\boldi}{{\bf i}}
\newcommand{\boldj}{{\bf j}}
\newcommand{\boldk}{{\bf k}}
\newcommand{\boldl}{{\bf l}}
\newcommand{\boldm}{{\bf m}}
\newcommand{\boldn}{{\bf n}}
\newcommand{\boldo}{{\bf o}}
\newcommand{\boldp}{{\bf p}}
\newcommand{\boldq}{{\bf q}}
\newcommand{\boldr}{{\bf r}}
\newcommand{\bolds}{{\bf s}}
\newcommand{\boldsp} {{\bf s}^{\prime}}
\newcommand{\boldt}{{\bf t}}
\newcommand{\boldu}{{\bf u}}
\newcommand{\boldv}{{\bf v}}
\newcommand{\boldw}{{\bf w}}
\newcommand{\boldx}{{\bf x}}
\newcommand{\boldy}{{\bf y}}
\newcommand{\boldz}{{\bf z}}

%BOLD LETTERS- CAPITAL LETTERS
\newcommand{\boldA}{{\bf A}}
\newcommand{\boldB}{{\bf B}}
\newcommand{\boldC}{{\bf C}}
\newcommand{\boldD}{{\bf D}}
\newcommand{\boldE}{{\bf E}}
\newcommand{\boldF}{{\bf F}}
\newcommand{\boldG}{{\bf G}}
\newcommand{\boldH}{{\bf H}}
\newcommand{\boldI}{{\bf I}}
\newcommand{\boldJ}{{\bf J}}
\newcommand{\boldK}{{\bf K}}
\newcommand{\boldL}{{\bf L}}
\newcommand{\boldM}{{\bf M}}
\newcommand{\boldN}{{\bf N}}
\newcommand{\boldO}{{\bf O}}
\newcommand{\boldP}{{\bf P}}
\newcommand{\boldQ}{{\bf Q}}
\newcommand{\boldR}{{\bf R}}
\newcommand{\boldS}{{\bf S}}
\newcommand{\boldT}{{\bf T}}
\newcommand{\boldU}{{\bf U}}
\newcommand{\boldV}{{\bf V}}
\newcommand{\boldW}{{\bf W}}
\newcommand{\boldX}{{\bf X}}
\newcommand{\boldY}{{\bf Y}}
\newcommand{\boldZ}{{\bf Z}}

%System Geometry 
\newcommand{\stx}{S_{\text{T}}}
\newcommand{\srx}{S_{\text{R}}}
\newcommand{\lt}{L_{\text{T}}}
\newcommand{\lr}{L_{\text{R}}}
\newcommand{\rhot}{\rho_{\text{T}}}
\newcommand{\phit}{\varphi_{\text{T}}}
\newcommand{\rhor}{\rho_{\text{R}}}
\newcommand{\phir}{\varphi_{\text{R}}}
\newcommand{\Pt}{P_{\text{T}}}
\renewcommand{\Pr}{P_{\text{R}}}
\newcommand{\Nt}{N_{\text{T}}}
\newcommand{\Nr}{N_{\text{R}}}
\newcommand{\Gt}{{\bf G_{\text{T}}}}
\newcommand{\Gr}{{\bf G_{\text{R}}}}
\newcommand{\lambdaA}{\lambda_{\boldA}}
\newcommand{\vA}{\mathbf{v}_{\mathbf{A}}}
\newcommand{\gammastar}{{\boldsymbol{\gamma}}^\star}
\newcommand{\Ndof}{N_{\text{DOF}}}
\newcommand{\bbeta}{{\bm {\beta}}}
\newcommand{\bchi}{{\bm{\chi}}}
\newcommand{\bsigma}{{\bm{\Sigma}}}
\newcommand{\blambda}{{\bm{\Lambda}}}
\newcommand{\invsigma}{{\bm{\Sigma}}^{-1}}
\newcommand{\pinvsigma}{{\bm{\Sigma}}^{\dagger}}
\newcommand{\tbw}{\tilde{\boldw}}
\newcommand{\tbsigma}{\tilde{\bm{\Sigma}}}
\newcommand{\hbbeta}{{\hat{\boldsymbol{\beta}}}}
\newcommand{\betan}{\beta_n}
\newcommand{\xin}{\xi_n}
\newcommand{\sumn}{\sum_{n=1}^\infty}
\newcommand{\tildexn}{\tilde{x}_n}
\newcommand{\xn}{x_n}
\newcommand{\xns}{x_n^\star}
\newcommand{\lambdas}{\lambda^\star}
\newcommand{\fz}{f_0}
\newcommand{\fo}{f_1}

\newcommand{\Ht}{\boldH_{\text{T}}}
\newcommand{\Hr}{\boldH_{\text{R}}}
\newcommand{\HE}{\boldH_{\text{E}}}

%---------------------%
%       References    %
%---------------------%
\newcommand{\fig}[1]{Fig.~\ref{#1}}
\newcommand{\sect}[1]{Sec.~\ref{#1}}
\newcommand{\apd}[1]{Appendix~\ref{#1}}
\newcommand{\eq}[1]{(\ref{#1})}

\newcommand{\gtnm}{g_{\text{T}, n, m}}
\newcommand{\Thetanm}{\boldsymbol{\Theta}_{n,m}}

\newcommand{\rt}{R_{\text{T}}}
\newcommand{\Frob}[1]{\left\lVert #1 \right\rVert_{\mathrm{F}}}
\newcommand{\Ree}{\operatorname{Re}}
\newcommand{\Imag}{\operatorname{Im}}
\newcommand{\modtwo}[1]{\operatorname{mod}_{2\pi}\left(#1\right)}
\maketitle
\acresetall

\section{Introduction}

\IEEEPARstart{T}{he} growing demand for high-throughput, low-latency wireless services is pushing current radio networks toward their physical limits, a trend expected to intensify with \ac{6G} systems, which will rely on higher carrier frequencies and larger, denser antenna apertures. Under these conditions, transmitter-receiver interactions often occur in the radiating near-field, where the plane-wave approximation no longer holds, and spherical wave propagation must be explicitly modeled. In this context, \ac{HMIMO} has emerged as a framework for shaping emitted and received \ac{EM} fields to fully exploit the channel \ac{DOF}. Prior studies have demonstrated substantial capacity gains even under \ac{LOS} propagation \cite{gong2023holographic}. However, practical \ac{HMIMO} hardware technologies are largely underdeveloped. Existing implementations have explored \acp{LIS} \cite{dardari2020communicating, decarli2021communication}, \acp{DMA} \cite{khan2026holistic}, and \ac{XL-MIMO} arrays \cite{torcolacci2024holographic} for communication and imaging/sensing, highlighting the potential of \ac{HMIMO} for \ac{ISAC} applications.

However, fully digital \ac{HMIMO} transceivers entail prohibitive hardware complexity and power consumption, exacerbating the \emph{digital bottleneck}. The \ac{ESP} paradigm \cite{dardari2024overview} offers an alternative path by shifting signal processing operations from the digital baseband to the \ac{EM} domain, thereby enabling direct wave manipulation during propagation, while drastically reducing the number of active components and energy-hungry \ac{RF} chains. Within this framework, \acp{SIM} have emerged as a compelling hardware platform to realize \ac{ESP}, as their cascaded programmable metasurface layers can synthesize complex wavefront transformations entirely in the wave domain.

Recently, \acp{SIM} have attracted significant attention owing to their broad applicability and promising wave-domain processing capabilities \cite{an2023stacked_holographic, liu2025stacked_Clerx, huang2024taskoriented}. A typical \ac{SIM} includes multiple stacked metasurface layers, each composed of hundreds of reconfigurable cells (meta-atoms), controlled by programmable electronics (e.g., \ac{FPGA}). Each meta-atom re-radiates toward subsequent layers following the Huygens-Fresnel principle, enabling dense EM-domain interaction. As a result, \acp{SIM} can emit and receive information-bearing \ac{EM} fields over almost the entire aperture, making them strong candidates for practical \ac{HMIMO} implementation. Compared to alternative \ac{HMIMO} technologies, \acp{SIM} offer higher spatial-domain gain \cite{AnXuetAl3:J23}, finer waveform control \cite{AnXuetAl2:J23, HassanetAl:J24}, and improved spatial multiplexing \cite{AnXuetAl2:J23, liu2024drl_multiuser}. Unlike conventional \acp{RIS}, which are typically deployed in the environment to improve coverage, \acp{SIM} are integrated directly into \ac{TX} and \ac{RX} devices. This integration can improve the link performance while reducing hardware complexity by limiting the number of active \ac{RF} chains and \ac{DAC}/\ac{ADC} units \cite{YaoetAl:J24, an2023multiuser}. Moreover, native wave-domain processing enables ultrafast, massive parallel operations, which are attractive for massive communications, sensing, and physical-layer security \cite{wang2024multi, huang2024taskoriented, niu2024physical_security}.  

In parallel, increasing attention has been devoted to the \ac{OAM} modes for \ac{HMIMO} transmissions \cite{lee2024continuous, dai2024optimal, jin2025achieving, vanwynsberghe2023walsh}. \ac{OAM} bases provide a low-complexity, structured mode set capable of supporting near-field \ac{LOS} spatial multiplexing, with limited performance loss relative to optimal designs based on \acp{CPSF}. Specifically, whereas CPSF-based optimal designs mandate computationally intensive eigenfunction solutions and accurate transmitter \ac{CSI}, \ac{OAM} schemes significantly simplify the transceiver architecture by eliminating the requirement for transmitter-side \ac{CSI} while retaining a significant number of orthogonal spatial channels \cite{TorDecDar:J23}. However, conventional \ac{OAM} generation often relies on hardware-specific devices, such as \acp{SPP} and \acp{UCA}, which are typically static, costly, or weakly reconfigurable \cite{chen2019orbital}. The multilayer programmability of \acp{SIM} offers a low-cost and reconfigurable platform for \ac{OAM}-enabled communications without specialized hardware. \ac{SIM}-\ac{OAM} integration is therefore promising, but still largely unexplored.

\subsection{State of the Art}
Research on \acp{SIM} has rapidly evolved across theories, modeling, and optimization. Surveys such as \cite{liu2025stacked_Clerx} have highlighted their potential for wave-domain processing in sensing, communications, and computing. The foundational work in \cite{an2023stacked_holographic} established \ac{SIM}-enabled \ac{HMIMO} communications via wave-domain precoding/combining, showing notable capacity gains over conventional \ac{MIMO} transmissions. Low-complexity architectures have also been proposed, such as meta-fiber designs, as in \cite{niu2025meta}.

Nevertheless, the optimization of \ac{SIM} configurations remains challenging owing to strong non-convexity. Alternating \ac{NMSE}-based methods \cite{an2023stacked_holographic} enable effective channel fitting but decouple amplitude and phase, potentially disrupting the phase correlations required by structured modes such as \ac{OAM}. For large-scale systems with statistical \ac{CSI}, joint transmit-covariance and \ac{SIM}-phase optimization have been proposed in \cite{papazafeiropoulos2024achievable}, whereas closed-form alternating updates for meta-fiber architectures reduce complexity \cite{niu2025meta}. 
Recently, \ac{AI}-driven methods have emerged as an alternative. Deep reinforcement learning has been applied to joint \ac{SIM} phase-shift and power allocation optimization in \ac{MU}-\ac{MISO} systems \cite{liu2024drl_multiuser}, and diffractive neural processing enables flexible amplitude-phase control \cite{darsena2025design_amplitude}. 
However, despite achieving significant gains in spectral efficiency, energy efficiency, or \ac{MSE} reduction, both conventional and \ac{AI}-driven approaches remain strictly anchored to standard performance metrics. Crucially, they do not explicitly enforce the helical phase structure required by \ac{OAM}, thereby failing to preserve the modal profile and spatial orthogonality that are essential for \ac{OAM}-based communications.

Within this context, \ac{OAM} integration into \ac{HMIMO}- and \ac{SIM}-based transmissions remains limited. Optical implementations have demonstrated \ac{OAM} multiplexing capabilities via metasurfaces \cite{fan2024multiplexed_oam}, but their extension to the \ac{RF} domain is hindered by discretization effects that can degrade modes' orthogonality. At \ac{RF}, continuous-aperture analyses have established the potential of near-field \ac{OAM} bases and hybrid Walsh-\ac{OAM} modes \cite{TorDecDar:J23, vanwynsberghe2023walsh}, but do not address discrete \ac{SIM} implementations and their optimization.

\subsection{Motivation and Contributions}

The above overview highlights that existing \ac{SIM} optimization frameworks, although effective for channel fitting or rate maximization, do not explicitly address the preservation of helical phase continuity and modal orthogonality and are thus not directly suited for \ac{OAM}-based systems. In particular, current methods typically decouple the amplitude and phase under \ac{NMSE}- or rate-based objectives \cite{an2023stacked_holographic, papazafeiropoulos2024achievable}, failing to preserve the helical phase correlations required for \ac{OAM} orthogonality. Moreover, existing frameworks lack dedicated metrics to quantify mode fidelity with respect to continuous-space references and  assess orthogonality under discrete implementations. Finally, both gradient-based and learning-based approaches \cite{an2023stacked_holographic, liu2024drl_multiuser} remain sensitive to initialization and local minima in highly non-convex phase landscapes, and do not provide a unified decoupled transmit/receive design strategy.

To address these limitations, we propose a correlation-driven optimization framework for \ac{OAM}-enabled \ac{SIM} systems. The main contributions of this study are as follows:

\begin{itemize}
\item We introduce a correlation-based objective that align generated \ac{EM} fields with ideal continuous-space \ac{OAM} references. Unlike \ac{NMSE}-based formulations, it enforces amplitude-weighted phase consistency and preserves the helical structure via complex-domain similarity, thereby improving robustness to low-amplitude regions. We further define dedicated metrics for the \ac{OAM} mode fidelity and inter-mode orthogonality in discrete \ac{SIM} architectures.

\item We develop a robust decoupled transmit/receive optimization strategy that combines multi-start gradient descent, adaptive step sizing, and layer-wise updates. The algorithm alternates exploration and refinement to mitigate local minima effects and enables end-to-end mode alignment without requiring full \ac{CSI}. Both fully optimized and hybrid configurations (i.e., one side ideal, one optimized) are considered to highlight the \ac{TX} and \ac{RX} side asymmetries induced by discretization.

\item We derive practical design rules from extensive simulations, linking metasurfaces' meta-atom density, number of layers, and link geometry to OAM mode fidelity and throughput. The results show that denser apertures significantly improve the per-mode correlation, additional layers enhance robustness to discretization but with diminishing returns beyond moderate depth, and transmit/receive asymmetries critically impact which side should be optimized in practice.

\end{itemize}

\subsection{Paper Organization}\label{sec:sections}

The remainder of this paper is organized as follows. Sec.~\ref{sec:OAM} reviews \ac{OAM} waves and the foundations of \ac{OAM}-based \ac{HMIMO}. Sec.~\ref{sec:SIMbasedOAM} presents the \ac{SIM} architecture and derives an end-to-end near-field system model. Sec.~\ref{sec:SIMoptimization} details the proposed \ac{SIM} optimization algorithm for \ac{OAM} transmission and reception. Sec.~\ref{sec:Results} presents the numerical analysis and related results. Sec.~\ref{sec:Conclusion} concludes the paper.

\subsection{Notation and Definitions}
Lowercase bold symbols (e.g., $\mathbf{x}$) denote vectors in \ac{3D} space, and uppercase bold symbols (e.g., $\mathbf{X}$) denote matrices. The identity matrix of size $N \times M$ is denoted by $\mathbf{I}_{N \times M}$, while the transpose and Hermitian operators are denoted by $(\cdot)^\top$, $(\cdot)^{\text{H}}$. The expectation and variance operators are $\E{\cdot}$ and $\Var{\cdot}$. The $\mathcal{L}_2$-norm of vector $\boldx$ is $\|\boldx\|$, and the Frobenius norm of matrix $\mathbf{X}$ is $\left\| \mathbf{X} \right\|_{\mathrm{F}}$; its trace is denoted by $\tr \left( \mathbf{X} \right)$. The imaginary unit is $\jmath$. The calligraphic symbols denote sets (e.g., $\mathcal{X}$). The notation $\mathbf{x} \sim \cn (\boldsymbol{\mu}, \boldsymbol{\Sigma})$ indicates a complex Gaussian random vector with mean $\boldsymbol{\mu}$ and covariance $\boldsymbol{\Sigma}$. The operator $\diag\left( \mathbf{x}\right)$ forms a diagonal matrix with the main diagonal $\mathbf{x}$, and $\operatorname{Re}\{\cdot\}$ denotes the real part.

\section{OAM-Based HMIMO Communications}\label{sec:OAM}

From a physical standpoint, \ac{EM} waves carrying non-zero \ac{OAM} exhibit helical wavefronts twisted around propagation axis\footnote{For this reason, an \ac{OAM} mode is frequently referred to as a ``vortex''.}. The helical structure is mathematically described by the exponential phase factor $e^{\jmath\ell \varphi}$, where $\ell \in \mathbb{Z}$ denotes the topological charge and $\varphi$ the transverse azimuthal angle defining the angular position on a plane orthogonal to the propagation direction. Each vortex is characterized by $\ell$, with $|\ell|$ indicating the number of phase twists per wavelength and the sign determining the twist chirality. The resulting vortex states associated with a given $\ell$ define the \ac{OAM} modes. Owing to their inherent orthogonality, \ac{OAM} modes enable spatial multiplexing, allowing multiple data streams to be simultaneously conveyed through distinct helical beams.

\begin{figure*}[t!]
    \centering
    \includegraphics[width=0.8\textwidth, keepaspectratio, trim=0 0 0 0, clip]{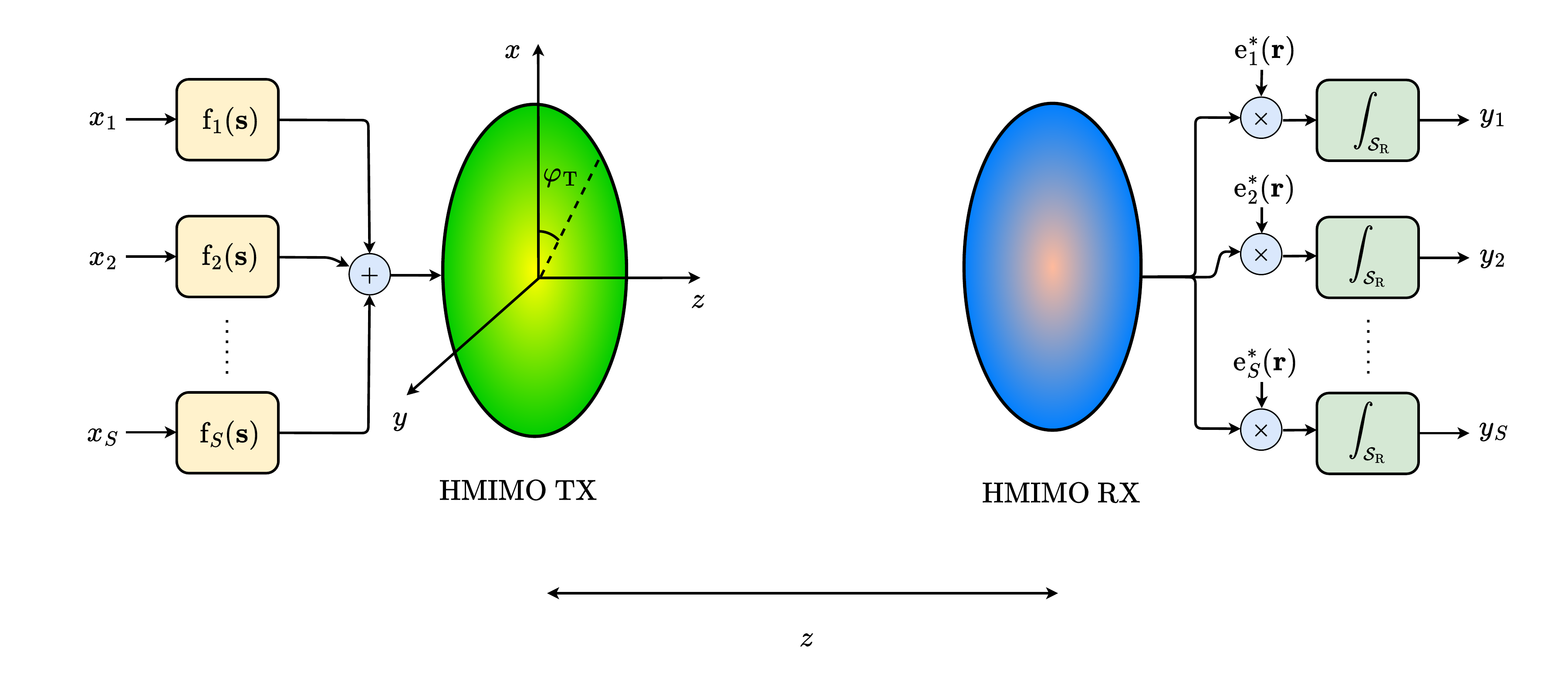}
    \caption{\ac{OAM}-based \ac{HMIMO} communication scheme in \ac{LOS}, with transmit surface coordinates $\bolds$ and receive surface coordinates~$\boldr$.}
    \label{fig:OAMScenario}
\end{figure*}

Abstracting now from the specific antenna geometries, consider a transmitting rectangular aperture with surface $S_{\text{T}}$ and area $A_{\text{T}} = \left| S_{\text{T}} \right|$, and a receiving rectangular aperture with surface $S_{\text{R}}$ and area $A_{\text{R}} = \left| S_{\text{R}} \right|$, as depicted in Fig.~\ref{fig:OAMScenario}. 

Let $\bolds \in S_{\text{T}}$ and $\boldr \in S_{\text{R}}$ denote the position vectors from the chosen reference origin to the points on the transmitting and receiving surfaces, respectively. A monochromatic source operating at frequency $f_0$, represented by $\mathrm{f}(\bolds)$, excites the transmitting aperture, resulting in a received \ac{EM} wave field $\mathrm{e}(\boldr)$ that satisfies the inhomogeneous Helmholtz equation \cite{dardari2024overview}, that is, 
\begin{equation}\label{eq:helmoltz}
\nabla^2 \mathrm{e}(\boldr) + \kappa^2 \mathrm{e}(\boldr) = -\mathrm{f}(\bolds)\,\,.
\end{equation}

Specifically, the solution to this equation under free-space propagation conditions is given by
\begin{equation}\label{eq:rxfunc}
\mathrm{e}(\boldr) = \int_{S_{\mathrm{T}}} G(\boldr, \bolds) \mathrm{f}(\bolds) \, \mathrm{d} \bolds\,\, ,
\end{equation}
where the scalar Green's function $ G(\boldr_1, \boldr_2) $ between points represented by vectors $ \boldr_1$ and $ \boldr_2 $ is given by
\begin{equation}\label{eq:Greenfunc}
G(\boldr_1, \boldr_2) = \frac{\exp \left(-\jmath \kappa \|\boldr_1 - \boldr_2\|\right)}{4 \pi \|\boldr_1 - \boldr_2\|}\,\, ,
\end{equation}
where $ \kappa = 2 \pi / \lambda $ is the free-space wavenumber and $ \lambda = c / f_0 $ represents the wavelength, where $ c $ is the speed of light. It is worth noting that \eqref{eq:Greenfunc} disregards the reactive field components, which become negligible at distances exceeding a few wavelengths, under the assumption of operation within the radiating near-field region. 

%----------------------------------------------------------------
%                     \ac{OAM} Scenario      
%----------------------------------------------------------------
%----------------------------------------------------------------

In this analysis, the \ac{EM} field is modeled as a complex-valued scalar quantity (e.g., under single-polarization conditions), although the physical field is vectorial. This simplification preserves the validity of the formulation, and can be straightforwardly extended to the full vector case.

Formally, communication modes can be represented through orthonormal basis expansions of the transmitting function $\mathrm{f}(\bolds)$ and the receiving function $\mathrm{e}(\boldr)$, denoted as $\mathrm{f}_i(\bolds)$ and $\mathrm{e}_i(\boldr)$, respectively, with $i = 1, 2, \ldots, \infty$. If the basis sets are designed such that a bijective correspondence exists between the $i$th transmitting basis function $\mathrm{f}_i(\bolds)$ and the $i$th receiving basis function $\mathrm{e}_i(\boldr)$, the presence of multiple communication modes is ensured \cite{Mil:C00}.

By expanding the wave functions $\mathrm{f}(\bolds)$ and $\mathrm{e}(\boldr)$ using the aforementioned orthonormal basis sets, which are complete, respectively, in $ S_{\text{T}}$ and $S_{\text{R}}$, we obtain
\begin{equation}
\begin{aligned}
\mathrm{f}(\bolds) &= \sum_{i=1}^{\infty} a_i \mathrm{f}_i(\bolds)\,\,, \\
\mathrm{e}(\boldr) &= \sum_{i=1}^{\infty} b_i \mathrm{e}_i(\boldr)\,\,,
\end{aligned}
\end{equation}
where $a_i$ and $b_i$ denote the corresponding expansion coefficients. Moreover, the orthonormality of such basis functions guarantees that, for $ i,j = 1,2,\ldots,\infty,\; i \neq j$, it holds
\begin{equation}
\begin{aligned}
\int_{S_{\text{T}}} \mathrm{f}_i(\bolds)\, \mathrm{f}_j^*(\bolds)\, \mathrm{d}\bolds &= \delta_{ij}\,, \\[2pt]
\int_{S_{\text{R}}} \mathrm{e}_i(\boldr)\, \mathrm{e}_j^*(\boldr)\, \mathrm{d}\boldr &= \delta_{ij}\,, 
\end{aligned}
\end{equation}

with $\delta_{ij}$ being the Kronecker delta function.

Although infinitely many orthonormal bases exist, whose computation requires solving a coupled eigenfunction problem \cite{dardari2024overview}, only a finite subset is strongly coupled and practically useful because of the finite size of the apertures. In this regard, \cite{TorDecDar:J23} introduced a simplified orthonormal transmit basis that exploits \ac{OAM} orthogonality and significantly reduces complexity relative to optimal \ac{CPSF} bases \cite{Xu:J17}. Indeed, while optimal eigen-bases strictly depend on the specific aperture geometry, yielding \acp{PSWF} for rectangular apertures and \ac{CPSF} for circular apertures, they are inherently distance-dependent, requiring full basis re-computation whenever the link distance changes. Conversely, the \ac{OAM} basis remains geometrically fixed and distance-invariant, decoupling the mode definition from the specific link range.

Under a paraxial propagation configuration, let $\boldp_{\text{T}} = (x_{\text{T}}, y_{\text{T}}) \in S_{\text{T}}$ and $\boldp_{\text{R}} = (x_{\text{R}}, y_{\text{R}}) \in S_{\text{R}}$ denote the Cartesian coordinates of the transmitting and receiving position vectors, respectively. By assuming a total number $S$ of well-coupled \ac{OAM}-based modes, the \ac{OAM}-carrying basis functions $ \left\{  \mathrm{f}_i(\boldp_{\text{T}}) \right\}_{i \in \mathcal{S}}$, with $\mathcal{S} = \{1,2, \ldots, S\}$, can be computed as \cite{TorDecDar:J23}
\begin{equation}
     \mathrm{f}_i(\boldp_{\text{T}}) = \frac{1}{ \sqrt{A_{\text{T}}}} \, e^{\jmath \ell_i \varphi_{\text{T}}}= \frac{1}{ \sqrt{A_{\text{T}}}} \, e^{\jmath \ell_i \arctan{\left(\frac{y_{\text{T}}}{x_{\text{T}}}\right)}} \,,
\end{equation}
where $\varphi_{\text{T}}$ denotes the transverse azimuthal angle on the transmitting surface antenna $S_{\text{T}}$, and 
\begin{align}
   \ell_i= 
        \begin{cases} -\frac{i-1}{2} & \mod(i,2)=1 \\
                      \frac{i}{2} &  \mod(i,2)=0
        \end{cases}
       , \;\; i \in \mathcal{S} \,.
\end{align}
Consequently, $\ell_i=0,+1,-1,+2,-2,\ldots, (S-1)/2$, assuming $S$ odd.

A critical characteristic of the \ac{OAM} modes is in their inherent beam divergence. It is well established that the divergence of \ac{EM} waves carrying \ac{OAM} increases with the topological charge, a phenomenon that is further exacerbated under near-field conditions \cite{xie2015performance}. Beam focusing techniques are commonly employed to counteract the resulting angular spread \cite{TorDecDar:J23}. These approaches introduce appropriate phase adjustments at the radiating aperture $S_{\text{T}}$ to compensate for propagation delays toward a designated focal point, thereby ensuring that the \ac{EM} field components combine coherently in that region. The outcome is the localized enhancement of the field intensity within a confined focal area. When operating under the Fresnel approximation \cite{Bal:B15}, such focusing on the receiving surface center can be accomplished through modified transmit \ac{OAM} basis functions, which can be computed as 
\begin{equation}\label{eq:OAM_foc_ideal}
     \mathrm{f}_i(\boldp_{\text{T}}) = \frac{1}{ \sqrt{A_{\text{T}}}} \, e^{\jmath \ell_i \arctan{\left(\frac{y_{\text{T}}}{x_{\text{T}}}\right)}}  e^{\jmath \kappa z \left( 1 + \frac{x_{\text{T}}^2 + y_{\text{T}}^2}{2 z^2}\right)}\,,
\end{equation}
where $z$ is the distance between the transmitting and receiving antenna centers.

Assume that $S$ data streams are transmitted simultaneously, carrying symbols $ \boldx = [ x_1, \ldots, x_i, \ldots, x_{S}]^\top \in \mathbb{C}^{S \times 1}$ and spatially multiplexed through the basis functions $ \mathrm{f}_i(\boldp_{\text{T}})$ in~\eqref{eq:OAM_foc_ideal}. The space-continuous transmitted signal is then
\begin{equation}
 x(\boldp_{\text{T}}) =  \sum_{i=1}^{S} x_i  \mathrm{f}_i(\boldp_{\text{T}})  \,,\;\; \boldp_{\text{T}} \in \mathcal{S}_{\text{T}} .
\end{equation}
Consequently, the information-bearing \ac{EM} field generated by mode superposition at distance $z$, i.e., over the receiving surface $S_{\text{R}}$, can be written in the absence of \ac{AWGN} as
\begin{equation}
 y(\boldp_{\text{R}}) = \sum_{i=1}^{S} y_i \mathrm{e}_i(\boldp_{\text{R}}) \,,\;\;\boldp_{\text{R}} \in \mathcal{S}_{\text{R}} \,,
\end{equation}
where $\{\mathrm{e}_i(\boldp_{\text{R}})\}, i \in \mathcal{S}$, denotes the \ac{OAM}-based receiving basis set. Specifically, these continuous receiving basis functions represent the \ac{EM} field distribution impinging on the receiving aperture after propagation through the \ac{HMIMO} channel, as described in \cite[Eq.~24]{TorDecDar:J23}.

%----------------------------------------------------------------
%                      \ac{SIM} Scenario      
%----------------------------------------------------------------
\begin{figure*}[t!]
    \centering
    \includegraphics[width=\textwidth, keepaspectratio, trim=0mm 0mm 0mm 0mm, clip]{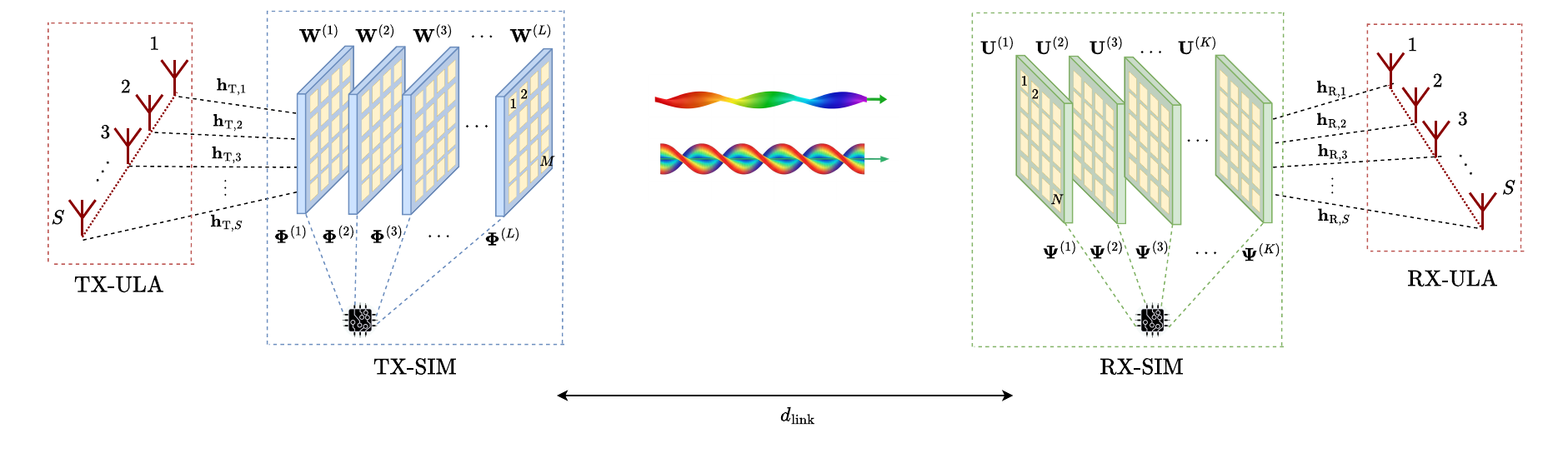}
    \caption{Illustration of a \ac{SIM}-aided \ac{LOS} \ac{HMIMO} scenario.}
    \label{fig:Scenario}
\end{figure*}
%----------------------------------------------------------------

\section{SIM-based HMIMO System}\label{sec:SIMbasedOAM}
The framework above is specialized for a specific \ac{HMIMO} antenna realization based on \acp{SIM}, which provides a practical implementation for manipulating \ac{EM} waves in the \ac{EM} wave domain. This enables the generation and shaping of structured wavefronts, including \ac{OAM} modes, directly over a wireless channel. Based on this perspective, the following section introduces the detailed \ac{SIM}-aided transceiver model for the considered \ac{LOS} \ac{HMIMO} system.

\subsection{System Model}\label{sec:SystemModel}

Consider the \ac{SIM}-aided \ac{HMIMO} system shown in Fig.~\ref{fig:Scenario}, where two identical \ac{SIM}-based transceivers operate in \ac{LOS} and paraxial configuration. The transmit and receive units are denoted by \ac{TX}-\ac{SIM} and \ac{RX}-\ac{SIM}, respectively.
The \ac{TX}-\ac{SIM} is illuminated by a \ac{ULA} with $S$ antenna elements, whereas the \ac{RX}-\ac{SIM} is followed by a receiving \ac{ULA} with $S$ elements. The same $S$ denotes the number of parallel data streams, that is, the orthogonal channels created at the \ac{EM} level.\footnote{A one-to-one correspondence is established between the $s$th antenna of the TX- and RX-\acp{ULA}, the $s$th transmitted symbol, and the $s$th \ac{OAM} mode, thereby defining $S$ parallel \ac{EM}-level channels for data transmission that harness the spatial \ac{DOF} of the near-field \ac{HMIMO} link.} Let $L$ and $K$ denote the number of metasurface layers at the \ac{TX}-\ac{SIM} and \ac{RX}-\ac{SIM}, respectively, with index sets $\mathcal{L} = \{1,2, \ldots, L\}$ and $\mathcal{K} = \{1,2, \ldots, K\}$. Each \ac{TX}-\ac{SIM} layer comprises $M$ meta-atoms ($M \geq S$), and each \ac{RX}-\ac{SIM} layer comprises $N$ meta-atoms ($N \geq S$), with index sets $\mathcal{M} = \{1,2, \ldots, M\}$ and $\mathcal{N} = \{1,2, \ldots, N\}$, respectively. Each layer is modeled as a uniform square lattice \cite{LiuetAl:J22}, i.e., a \ac{UPA} at both the \ac{TX} and \ac{RX}. At the \ac{TX}-\ac{SIM}, the intra-meta-atom spacing, meta-atom area, and intra-layer spacing are denoted by $d_{\mathrm{T}, \text{atom}}$, $A_{\mathrm{T},\text{atom}}$, and $d_{\mathrm{T},\text{layer}}$. The corresponding \ac{RX}-side quantities are $d_{\mathrm{R}, \text{atom}}$, $A_{\mathrm{R}, \text{atom}}$, and $d_{\mathrm{R},\text{layer}}$.

Each meta-atom at both \ac{TX}- and \ac{RX}-\ac{SIM} is assumed to apply phase-only control to the incident \ac{EM} wave without amplitude modulation. Furthermore, the meta-atoms are assumed to be uncoupled, e.g., by maintaining an inter-element spacing of $\lambda/2$, thereby ensuring that the \ac{SIM} response can be mathematically modeled via a diagonal transmission matrix.
For simplicity, each metasurface layer is modeled as a transmissive surface that captures the incident signal on the side facing the \ac{TX}-\ac{ULA} and re-radiates it toward the next layer with element-wise phase shifts. All the layers in both \acp{SIM} are assumed passive and lossless, i.e., they redistribute the incident power without amplification. Let $\phi_m^{(l)}=e^{\jmath \theta_m^{(l)}}$ denote the transmission coefficient of the $m$th meta-atom on the $l$th layer, with $\theta_m^{(l)} \in[0,2 \pi),\, m \in \mathcal{M},\, l \in \mathcal{L}$. The phase-shift matrix of layer $l$ is
$\boldsymbol{\Phi}^{(l)}=\diag\left(\phi_1^{(l)}, \phi_2^{(l)}, \cdots, \phi_M^{(l)}\right) \in \mathbb{C}^{M \times M}$, with $\left(\boldsymbol{\Phi}^{(l)}\right)^H \boldsymbol{\Phi}^{(l)} = \boldI_M,\, l \in \mathcal{L}$. Similarly, $\psi_n^{(k)}=e^{\jmath \xi_n^{(k)}}$ denotes the phase shift applied by the $n$th meta-atom on the $k$th receiver layer, with $\xi_n^{(k)} \in[0,2 \pi), n \in \mathcal{N}, k \in \mathcal{K}$. The corresponding transmission-coefficient matrix is $\boldsymbol{\Psi}^{(k)}=\diag\left(\psi_1^{(k)}, \psi_2^{(k)}, \cdots, \psi_N^{(k)}\right) \in \mathbb{C}^{N \times N}$, such that $\left(\boldsymbol{\Psi}^{(k)}\right)^H \boldsymbol{\Psi}^{(k)} = \boldI_N,\, k \in \mathcal{K}$.

Furthermore, let $\mathbf{W}^{(l)} \in \mathbb{C}^{M \times M}$, $l \in \mathcal{L} \setminus \{1\}$, denote the \ac{TX}-\ac{SIM} intra-layer propagation matrix from layer $(l-1)$ to layer $l$. According to Rayleigh-Sommerfeld diffraction theory \cite{Bal:B15}, the $(m,m^{\prime})$th entry of $\mathbf{W}^{(l)}$ is
\begin{equation}\label{eq:w_mml}
w_{m, m^{\prime}}^{(l)}=\frac{ A_{\mathrm{T},\text{atom}} \, \cos\left(\chi_{m, m^{\prime}}^{(l)}\right)}{d_{m, m^{\prime}}^{(l)}}\left(\frac{1}{2 \pi d_{m, m^{\prime}}^{(l)}}-\frac{j}{\lambda} \right) e^{\jmath \kappa d_{m, m^{\prime}}^{(l)}}\,,
\end{equation}
where $\kappa=2\pi/\lambda$ is the wavenumber, $\lambda$ is the wavelength, $d_{m, m^{\prime}}^{(l)}$ is the distance from the $m^{\prime}$th meta-atom of layer $(l-1)$ to the $m$th meta-atom of layer $l$, and $\chi_{m, m^{\prime}}^{(l)}$ is the angle between the propagation direction and the normal to layer $(l-1)$. The overall \ac{TX}-\ac{SIM} response is
\begin{equation}
\mathbf{P}=\boldsymbol{\Phi}^{(L)} \mathbf{W}^{(L)} \ldots \boldsymbol{\Phi}^{(2)} \mathbf{W}^{(2)} \boldsymbol{\Phi}^{(1)} \mathbf{W}^{(1)} \in \mathbb{C}^{M \times M}.
\end{equation}
Similarly, let $\mathbf{U}^{(k)} \in \mathbb{C}^{N \times N}$, $k \in \mathcal{K}/\{1\}$, denote the propagation between receive layers $(k-1)$ and $k$. The $(k,k^{\prime})$ element of $\mathbf{U}^{(k)}$ is
\begin{equation}\label{eq:u_nn_k}
u_{n,n^{\prime}}^{(k)}=\frac{A_{\mathrm{R},\text{atom}}\,  \cos \left(\zeta_{n,n^{\prime}}^{(k)}\right)}{d_{n,n^{\prime}}^{(k)}}\left(\frac{1}{2 \pi d_{n,n^{\prime}}^{(k)}}- \frac{j}{\lambda}\right) e^{\jmath \kappa d_{n,n^{\prime}}^{(k)} }\,,
\end{equation}
where $d_{n,n^{\prime}}^{(k)}$ is the distance from the $n^{\prime}$th meta-atom of receive layer $(k-1)$ to the $n$th meta-atom of layer $k$, and $\zeta_{n,n^{\prime}}^{(k)}$ is the angle between the propagation direction and the normal direction of layer $(k-1)$. The overall \ac{RX}-\ac{SIM} response is as follows
\begin{equation}
\mathbf{Q}=\boldsymbol{\Psi}^{(K)} \mathbf{U}^{(K)} \ldots \boldsymbol{\Psi}^{(2)} \mathbf{U}^{(2)} \boldsymbol{\Psi}^{(1)} \mathbf{U}^{(1)} \in \mathbb{C}^{N \times N}\, .
\end{equation}

The coupling between the $s$th transmit \ac{ULA} element and the first \ac{TX}-\ac{SIM} layer is described by $\mathbf{h}_{\mathrm{T}, s} \in \mathbb{C}^{M\times 1}, \, s \in \mathcal{S}$. Its entries are obtained by replacing $d_{m, m^{\prime}}^{(l)}$ in \eqref{eq:w_mml} with the distance between the $s$th \ac{ULA} element and the $m$th meta-atom of the first \ac{TX}-\ac{SIM} layer.\footnote{For simplicity, both \ac{TX} and \ac{RX} \acp{ULA} are modeled with isotropic antenna elements, which ensures that the assessment captures the intrinsic multiplexing capabilities of the \ac{SIM}-aided transceiver while preserving analytical tractability. The effect of realistic antenna patterns will be addressed in future research.} The corresponding coupling matrix is $\Ht =\left[\mathbf{h}_{\mathrm{T}, 1} ,\, \ldots,\,\mathbf{h}_{\mathrm{T}, s},\, \ldots,\, \mathbf{h}_{\mathrm{T}, S}\right] \in \mathbb{C}^{M \times S}$.

Similarly, the coupling vector between the last \ac{RX}-\ac{SIM} layer and the $s$th receive \ac{ULA} element is $\mathbf{h}_{\mathrm{R}, s}^\top \in \mathbb{C}^{1 \times N},\, s \in \mathcal{S}$. It is computed by replacing $d_{n, n^{\prime}}^{(k)}$ in \eqref{eq:u_nn_k} with the distance between the $n$th meta-atom of the last \ac{RX}-\ac{SIM} layer and the $s$th receive antenna. The resulting coupling matrix is $\Hr =\left[\mathbf{h}_{\mathrm{R}, 1}^\top,\, \ldots, \, \mathbf{h}_{\mathrm{R}, s}^\top,\, \ldots, \,\mathbf{h}_{\mathrm{R}, S}^\top\right]^\top \in \mathbb{C}^{S \times N}$. As a consequence, it holds $\mathbf{W}^{(1)} = \boldI_{M \times M}$ and $\mathbf{U}^{(1)} = \boldI_{N \times N}$. 
Moreover, the transmit and receive \acp{ULA} are assumed to be placed close to their corresponding \acp{SIM}, i.e., within their near-field regions.

\subsection{Near-Field \ac{HMIMO} Channel Model}\label{sec:HMIMOchannel}

The \ac{HMIMO} channel between the \ac{TX}- and \ac{RX}-\ac{SIM} is modeled under radiative near-field (Fresnel) conditions, such that it satisfies
\begin{equation}
0.62\, \sqrt{\frac{D^3}{\lambda}}\le d_{\text{link}} \le \frac{2\, D^2}{\lambda} \, ,
\end{equation}
where $d_{\text{link}}$ denotes the separation between the \ac{TX}-\ac{SIM} and \ac{RX}-\ac{SIM}, and $D$ represents the effective aperture size, defined by the largest dimension between the TX and RX \ac{SIM} layers facing the \ac{HMIMO} channel, assumed to be equal here for simplicity~\cite{Bal:B15}.

The near-field \ac{LOS} \ac{HMIMO} channel from the output surface of the \ac{TX}-\ac{SIM} to the input surface of the \ac{RX}-\ac{SIM} is denoted by $\boldG =\left\{  g_{n,m}  \right\} \in \mathbb{C}^{N \times M}$, with
\begin{equation}
     g_{n,m} = \frac{\lambda}{4\, \pi\, d_{n,m}} \, e^{- j \frac{2\, \pi}{\lambda} \,d_{n,m} }\,, \,\, m \in \mathcal{M},\, n \in \mathcal{N} \,, \\
\end{equation}
where $d_{n,m}$ is the distance between the $m$th \ac{TX}-\ac{SIM} meta-atom in layer $L$ and the $n$th \ac{RX}-\ac{SIM} meta-atom in layer $1$. This model uses exact pairwise geometry and therefore captures near-field propagation effects.

The transmit \ac{ULA} emits $ \boldx = [ x_1, \ldots, x_s, \ldots, x_{S}]^\top \in \mathbb{C}^{S \times 1}$ with $||\boldx||^2 \leq \Pt$, where $\Pt$ is the available transmit power.
After propagation through the \ac{TX}-\ac{SIM}, \ac{LOS} \ac{HMIMO} channel, and \ac{RX}-\ac{SIM}, the received signal at the \ac{ULA} is
\begin{align}
    \boldy = \HE \, \boldx + \tilde{\boldw}  =  \Hr \boldQ \boldG \boldP \Ht \boldx + \tilde{\boldw} \, ,
\end{align}
where $ \HE \in \mathbb{C}^{S\times S} $ is the end-to-end effective channel from the transmitting to receiving \ac{ULA}. The noise term is $\tilde{\boldw}= \Hr \boldQ \boldw\in \mathbb{C}^{S\times 1}$, where $\boldw \in \mathbb{C}^{N\times 1} \sim \cn \left(0, \sigma_{\mathrm{w}}^2\boldI_{S}\right)$ is the input \ac{AWGN} vector at the \ac{RX}-\ac{SIM}, with \ac{i.i.d.} complex Gaussian entries and variance $\sigma_{\mathrm{w}}^2$. Therefore, $\tilde{\boldw} \sim \cn \left(0,\Sigma_{\tilde{\mathrm{w}}}\boldI_{S}\right)$, where $ \Sigma_{\tilde{\mathrm{w}}} = \Var{\tilde{\boldw}} = \tr \left( \boldC \right)$ and 
\begin{align}
    \boldC =  \mathbb{E}\left[\tilde{\boldw} \tilde{\boldw}^H\right] 
= \sigma_{\mathrm{w}}^2 \Hr \boldQ \boldQ^H \Hr^H \,\in \mathbb{C}^{S\times S}\,.
\end{align}

\subsection{Target OAM Discrete Basis Functions}\label{Sec:OAM_HMIMO_bases}

We now focus on \ac{OAM} transmission and reception using \acp{SIM}. Prior \ac{OAM}-\ac{HMIMO} studies (e.g., \cite{lee2024continuous, jin2025achieving, TorDecDar:J23}) usually assumed continuous antenna apertures, whereas practical systems are discrete. In the proposed architecture, each meta-atom acts as a spatial sample; therefore, the transmitted and received \ac{EM} fields are represented by vectors sampled on metasurface elements. This discretization captures the finite spatial resolution while enabling accurate \ac{OAM} mode synthesis and detection.
In the following, we leverage the focused continuous-space \ac{OAM} bases in \eqref{eq:OAM_foc_ideal} to reduce the beam divergence and improve the energy transfer between \ac{TX}- and \ac{RX}-\acp{SIM}. 

Specifically, for a given vector of transmitted symbols $\boldx$, let $\boldr = [r_1, \ldots, r_m, \ldots, r_M]^\top \in \mathbb{C}^{M \times 1}$ denote the ideal \ac{OAM}-encoded target signal that the \ac{TX}-\ac{SIM} aims to synthesize at its final layer $L$, with entries
\begin{equation}
 r_m =  \sum_{s=1}^{S} x_s  \mathrm{f}_s\left(\boldp_m^{(L)}\right) = \sum_{s=1}^{S} x_s \mathrm{f}_{ms}  \,\,, \;\;m\in \mathcal{M},
\end{equation}
where
\begin{equation}\label{eq:f_basis}
    \mathrm{f}_{ms}  = \frac{1}{ \sqrt{A_{\text{SIM}, \mathrm{T}}}} \, e^{\jmath \ell_s \arctan{\left(\frac{y_{m}^{(L)}}{x_{m}^{(L)}}\right)}}  e^{\jmath \kappa d_{\text{c},m} \left(\! 1 + \frac{\left(x_{m}^{(L)}\right)^2 + \left(y_{m}^{(L)}\right)^2}{2 d_{\text{c},m}^2}\right)}
\end{equation}
represents the $s$th \ac{OAM} transmit basis function in $\left\{\mathrm{f}_{ms} \right\}, \, s \in \mathcal{S}$, associated with mode order $\ell_s$. Here, $\boldp_m^{(L)}=\left(x_{m}^{(L)}, y_{m}^{(L)}\right)$ are the Cartesian coordinates of the $m$th meta-atom on the \ac{TX}-\ac{SIM} layer $L$, and $d_{\text{c},m}$ is the distance between that the meta-atom and the center of the \ac{RX}-\ac{SIM} aperture. 
Accordingly, let $\boldf_s \in \mathbb{C}^{M \times 1}\,, s \in \mathcal{S},$ denote the precoding vector of mode $s$. Stacking these vectors gives $\boldF = \left[ \boldf_1\,,\ldots\,,\boldf_s\,,\ldots\,,\boldf_S \right] = \left\{ \mathrm{f}_{ms}\right\} \in \mathbb{C}^{M \times S}$, which defines the desired \ac{EM}-domain mapping from the $S$ symbols to the focused \ac{OAM} basis functions. Therefore, $\boldF$ represents the target precoding map to be synthesized by the cascade $\boldP\Ht$.

Similarly, let $\boldv = [ v_1, \ldots, v_n, \ldots v_N]^\top\in \mathbb{C}^{N \times 1}$ denote the ideal \ac{OAM}-carrying target signal impinging on the first \ac{RX}-\ac{SIM} layer, the elements of which are given by
\begin{equation}\label{eq:e_basis}
   v_n = \sum_{s=1}^{S} y_s \mathrm{e}_s\left(\boldp_n^{(1)}\right) =\sum_{s=1}^{S} y_s \mathrm{e}_{sn} \,\,,\;\; n \in \mathcal{N},
\end{equation}
where $\left\{\mathrm{e}_{sn} \right\}_{s \in \mathcal{S}}$ represents the \ac{OAM}-based receive basis evaluated at the $n$th meta-atom of the first \ac{RX}-\ac{SIM} layer with coordinates $\boldp_n^{(1)}=\left(x_{n}^{(1)}, y_{n}^{(1)}\right)$. Specifically, these discrete basis functions are obtained by evaluating at each meta-atom the continuous \ac{EM} field distribution generated by transmitting the ideal \ac{OAM} modes in \eqref{eq:OAM_foc_ideal} and propagating them through the \ac{HMIMO} channel onto the \ac{RX}-\ac{SIM} surface, as characterized in \cite[Eq.~24]{TorDecDar:J23}.
Accordingly, an ideal combining matrix $\mathbf{E} = [ \bolde_1^\top, \,\dots, \, \bolde_s^\top, \,\dots, \,\bolde_S^\top ]^\top \in \mathbb{C}^{S \times N}$ can be defined at the \ac{RX}-\ac{SIM}, where each row $\bolde_s^\top \in \mathbb{C}^{1 \times N}\,, s \in \mathcal{S}$, denotes the combining vector employed to extract the $s$th data stream for symbol decoding.

For symbol recovery, we adopt \ac{MF} detection \cite{TorDecDar:J23}. Given this choice, the target combining matrix is $\boldE = \left\{ \mathrm{e}_{sn}^* \right\} \in \mathbb{C}^{S \times N}$, where each entry is the complex conjugate of the corresponding discretized receive basis function in \eqref{eq:e_basis}. Specifically, to recover symbol $x_s$ with maximum \ac{SNR}, $\boldQ$ and $\Hr$ must jointly synthesize the response $\Hr \boldQ$ consistent with the target matrix $\boldE$.

Therefore, matrices $\boldF$ and $\boldE$ serve as references for the optimization stage, which quantifies how closely practical \ac{SIM} configurations reproduce these mappings on both the \ac{TX} and \ac{RX} sides.

\section{SIM Configuration Optimization}\label{sec:SIMoptimization}

Under these considerations, the optimization task can be formalized as two distinct, structurally analogous problems. The \ac{TX}-\ac{SIM} configuration design is obtained by solving 
\begin{align}
\mathcal{P}_{\mathrm{T}}\!: \,\underset{\left\{\theta_m^{(l)}\right\}}{\operatorname{minimize}}& \; 
\begin{aligned}[t]
\mathcal{L}_{\mathrm{T}}
&= \frac{1}{2} \!\left(\!1 - \frac{
\operatorname{Re}\left\{\operatorname{tr}\left(\boldF^{\text{H}} \boldP \Ht\right)\right\}
}{
\|\boldF\|_{\mathrm{F}} \| \boldP \Ht \|_{\mathrm{F}}
} \right)
\end{aligned} \label{eq:Pt_cost_func}\\
\text{s.t.} \quad & 
\boldP = \boldsymbol{\Phi}^{(L)} \mathbf{W}^{(L)} \cdots \boldsymbol{\Phi}^{(1)} \mathbf{W}^{(1)} \\
& \boldsymbol{\Phi}^{(l)} = \diag\left(e^{\jmath \theta_1^{(l)}}, \ldots, e^{\jmath \theta_M^{(l)}}\right), \, l \in \mathcal{L} \\
& \theta_m^{(l)} \in [0, 2\pi), \!m \in \mathcal{M}, \; l \in \mathcal{L} \,.
\end{align}
Analogously, the \ac{RX}-\ac{SIM} design is
\begin{align}
\mathcal{P}_{\mathrm{R}}\!:\! \underset{\{\xi_n^{(k)}\}}{\operatorname{minimize}}& \; 
\begin{aligned}[t]
\mathcal{L}_{\mathrm{R}}
&= \frac{1}{2} \!\left(\!1 - \frac{
\operatorname{Re}\!\left\{\operatorname{tr}\left(\mathbf{E}^{\text{H}} \mathbf{H}_{\mathrm{R}} \mathbf{Q}\right)\right\}
}{
\|\mathbf{E}\|_{\mathrm{F}} \| \mathbf{H}_{\mathrm{R}} \mathbf{Q} \|_{\mathrm{F}}
}\!\right)
\end{aligned} \label{eq:Pr_cost_func}\\
\text{s.t.} \quad & 
\mathbf{Q}=\boldsymbol{\Psi}^{(K)} \mathbf{U}^{(K)} \ldots \boldsymbol{\Psi}^{(1)} \mathbf{U}^{(1)}  \\
& \boldsymbol{\Psi}^{(k)}\! = \operatorname{diag}\! \left(e^{\jmath \xi_1^{(k)}}, \cdots, e^{\jmath \xi_N^{(k)}}\right)\!, \, k \in \mathcal{K} \\
& \xi_n^{(k)} \in [0, 2\pi), \ \! n \in \mathcal{N}, \; k \in \mathcal{K} 
\end{align}
The correlation-based objectives in \eqref{eq:Pt_cost_func} and \eqref{eq:Pr_cost_func} depart from the conventional \ac{NMSE}-based formulations by directly maximizing the normalized complex correlation~\cite{rudin1987real}. Unlike \ac{NMSE}, this criterion enables amplitude-weighted phase alignment, naturally preserving the helical phase structure of the \ac{OAM} modes even in the presence of amplitude nulls and phase singularities. In our preliminary evaluations, replacing the proposed objective with an \ac{NMSE} criterion consistently led to poor convergence, yielding low-\ac{SINR} solutions with reduced inter-mode orthogonality and phase distributions that no longer matched the desired \ac{OAM} profiles. This behavior is consistent with the known limitations of \ac{NMSE}-based objectives for phase-sensitive wavefront synthesis~\cite{an2023stacked_holographic}.

The resulting optimization problems are highly non-convex. To address this, we adopt a multi-start gradient-based strategy using \ac{Adam}, followed by a refinement stage to improve convergence stability and mitigate poor local minima, thus alternating exploration and fine-tuning. Details are presented in Algorithm~\ref{alg:SIMopt}. Notably, because $\mathcal{P}_{\mathrm{T}}$ and $\mathcal{P}_{\mathrm{R}}$ share the same structural form and are solved independently, the proposed framework enables decoupled TX/RX optimization without requiring explicit \ac{CSI} exchange or joint transceiver coordination.

%%%%%%%%%%%%%%%%%%%%%%%%%%%%%%%%%%%%%%%%%%%%%%%%%%%%
\begin{algorithm}[t!]
\small
\SetAlgoLined % Mostra le linee verticali di blocco
\SetInd{0.6em}{0.6em}
\SetKwRepeat{Do}{do}{while}
\caption{\ac{SIM} optimization for \ac{OAM} mode generation}\label{alg:SIMopt}

% ================= INPUT =================
\BlankLine
\KwIn{$\mathbf{F} \in \mathbb{C}^{M \times S}$, $\mathbf{H}_{\text{T}} \in \mathbb{C}^{M \times S}$, $\{\mathbf{W}^{(l)}\}_{l\in \mathcal{L}}$, $N_{\text{runs}}$, $N_{\text{scr}}$, $N_{\text{ref}}$, $\eta_{\text{scr}}$, $\eta_{\text{ref}}$}

% ================= GLOBAL INIT =================
\BlankLine
\underline{\textbf{Global Initialization}:} 
$\mathcal{L}^{\star} \gets +\infty$,\, $\boldsymbol{\Theta}^{\star} \gets \text{null}$,\, $\mathbf{P}^{\star} \gets \text{null}$ \;

% ================= MULTI-START SCREENING =================
\BlankLine
\underline{\textbf{Multi-start Screening Phase}:}\,
\For{$r \leftarrow 1$ \KwTo $N_{\text{runs}}$}{
    \BlankLine
    \textbf{Run Initialization:}\;
    $\theta_m^{(l)} \sim \mathcal{U}[0, 2\pi)$, $\forall m \in \mathcal{M}, \; l \in \mathcal{L}$, \;
    $\mathbf{m}^{(l)} \gets \mathbf{0}_M$, $\mathbf{v}^{(l)} \gets \mathbf{0}_M$, $\forall l$ \;
    
    \For{$t \leftarrow 1$ \KwTo $N_{\text{scr}}$}{
        \textbf{(\emph{i}) Forward Pass:}\;
        $\mathbf{P} = \boldsymbol{\Phi}^{(L)} \mathbf{W}^{(L)} \cdots \boldsymbol{\Phi}^{(1)} \mathbf{W}^{(1)}$\;
        $\mathbf{A} \gets \mathbf{P} \mathbf{H}_{\text{T}}$\;
        $\mathcal{L}_{\mathrm{T}} = \tfrac{1}{2}\Big( 1 - \operatorname{Re}\{ \text{tr}( \mathbf{F}^{\text{H}} \mathbf{A} ) \}/( \lVert \mathbf{F} \rVert_{\mathrm{F}} \lVert \mathbf{A} \rVert_{\mathrm{F}} ) \Big)$ \;
        
        \BlankLine
        \textbf{(\emph{ii}) Backward Pass:}\;
        Compute phase gradients $\nabla_{\boldsymbol{\theta}} \mathcal{L}_{\mathrm{T}}$ \;
        
        \BlankLine
        \textbf{(\emph{iii}) Update Step:}\;
        $\boldsymbol{\theta}^{(l)} \gets \text{Adam}(\boldsymbol{\theta}^{(l)}, \nabla_{\boldsymbol{\theta}^{(l)}} \mathcal{L}_{\mathrm{T}}, \mathbf{m}^{(l)}, \mathbf{v}^{(l)}, \eta_{\text{scr}}, t)$\;
        $\boldsymbol{\theta}^{(l)} \gets \text{mod}_{2\pi}\left(\boldsymbol{\theta}^{(l)}\right)$ \;
    }
    
    \BlankLine
    \textbf{Run Evaluation:}\;
    \If{$\mathcal{L}_{\mathrm{T}} < \mathcal{L}^{\star}$}{
        $\mathcal{L}^{\star} \gets \mathcal{L}_{\mathrm{T}}$, $\boldsymbol{\Theta}^{\star} \gets \left\{\boldsymbol{\theta}^{(l)}\right\}_{l\in \mathcal{L}}$\;
    }
}

% ================= MULTI-PHASE REFINEMENT =================
\BlankLine
\underline{\textbf{Multi-Phase Refinement}:} 
$\boldsymbol{\Theta}_{\text{ref}} \gets \boldsymbol{\Theta}^{\star}$, $\eta_{\text{base}} \gets \eta_{\text{ref}}$\;

\For{$p \leftarrow 1$ \KwTo $N_{\text{phases}}$}{
    $\mathbf{m}^{(l)} \gets \mathbf{0}_M$, $\mathbf{v}^{(l)} \gets \mathbf{0}_M$, $\forall l$ \;
    $\mathcal{L}_{\text{start}} \gets \mathcal{L}^{\star}$\;
    
    \For{$t \leftarrow 1$ \KwTo $N_{\text{ref}}$}{
        $\eta_t \gets \eta_{\text{min}} + \tfrac{1}{2}(\eta_{\text{base}} - \eta_{\text{min}})(1 + \cos(\pi t / N_{\text{ref}}))$ \;
        Repeat forward/backward/Adam update with $(\boldsymbol{\Theta}_{\text{ref}}, \eta_t)$\;
        
        \If{$\mathcal{L}_{\mathrm{T}} < \mathcal{L}^{\star}$}{
            $\mathcal{L}^{\star} \gets \mathcal{L}_{\mathrm{T}}$, $\boldsymbol{\Theta}^{\star} \gets \boldsymbol{\Theta}_{\text{ref}}$\;
        }
    }
    
    $\Delta_{\text{rel}} \gets (\mathcal{L}_{\text{start}} - \mathcal{L}^{\star}) / \max(\mathcal{L}_{\text{start}}, 10^{-12})$\;
    \If{$\Delta_{\text{rel}} < \tau_{\text{conv}}$}{
        \textbf{break}\;
    }
    $\eta_{\text{base}} \gets \gamma \,\eta_{\text{base}}$ \;
}
% ================= OUTPUT =================
\BlankLine
\KwOut{$\mathbf{P}^{\star} = \boldsymbol{\Phi}^{(L)} \mathbf{W}^{(L)} \cdots \boldsymbol{\Phi}^{(1)} \mathbf{W}^{(1)}$,\, $\boldsymbol{\Theta}^{\star}$, \,$\mathcal{L}^{\star}$} 
\end{algorithm}
%%%%%%%%%%%%%%%%%%%%%%%%%%%%%%%%%%%%%%%%%%%%%%%%%%%%

Particularly, $N_{\text{runs}}$ independent phase configurations $\boldsymbol{\Theta}$ are drawn uniformly over $[0,2\pi)$ (multi-start initialization). For each run, $N_{\text{scr}}$ gradient-based iterations are performed. At each iteration, a forward pass constructs the \ac{SIM} configuration matrix $\boldP$ by cascading the intra-layer propagation matrices $\{\mathbf{W}^{(l)}\}$ and phase-shift operators $\{\boldsymbol{\Phi}^{(l)}\}$, and computes the effective precoder $\mathbf{A}_{\text{T}} = \boldP \Ht$. The loss $\mathcal{L}_{\mathrm{T}}$ is then evaluated using the normalized correlation in \eqref{eq:Pt_cost_func}.

The gradients with respect to the phase variables $\{\boldsymbol{\theta}^{(l)}\}$ are computed via backpropagation using Wirtinger calculus, accounting for the cascaded structure of the metasurface layers. The resulting gradients are used to update the phases through the \ac{Adam} optimizer, followed by projection onto the $[0,2\pi)$ range.

After the screening phase, the best initialization is selected and refined through $N_{\text{phases}}$ additional phase optimization stages. Each stage consists of $N_{\text{ref}}$ iterations with a progressively decreasing learning rate following a cosine annealing schedule \cite{loshchilov2017sgdr}, improving the convergence toward a local minimum. Early stopping is applied based on the relative loss improvement.

Lastly, the optimization algorithm returns the set of phase shifts $\boldsymbol{\Theta}^\star$, the corresponding \ac{TX}-\ac{SIM} transformation $\boldP^\star$, and the achieved loss $\mathcal{L}^\star$. As a consequence, the resulting transmitter-side processing is described by the effective wave-domain precoder $\boldA_{\mathrm{T}} = \boldP^{\star} \Ht \in \mathbb{C}^{M \times S}$.

Dually, the \ac{RX}-\ac{SIM} optimization follows the same procedure, yielding optimized phases $\boldsymbol{\Xi}^\star$, transformation $\boldQ^\star$, and loss $\mathcal{L}_{\mathrm{R}}^\star$. This leads to the effective combining operator $\boldA_{\mathrm{R}} = \Hr \boldQ^{\star} \in \mathbb{C}^{S \times N}$.

\vspace{12mm}

\section{Numerical Results}\label{sec:Results}

\subsection{Simulation Setup}\label{sec:sim_setup}

%------ FIGURE: corr_vs_L (double column) ------
\begin{figure*}[t!]
    \centering
    \begin{minipage}[t]{0.42\textwidth}
        \centering
        \includegraphics[width=\linewidth]{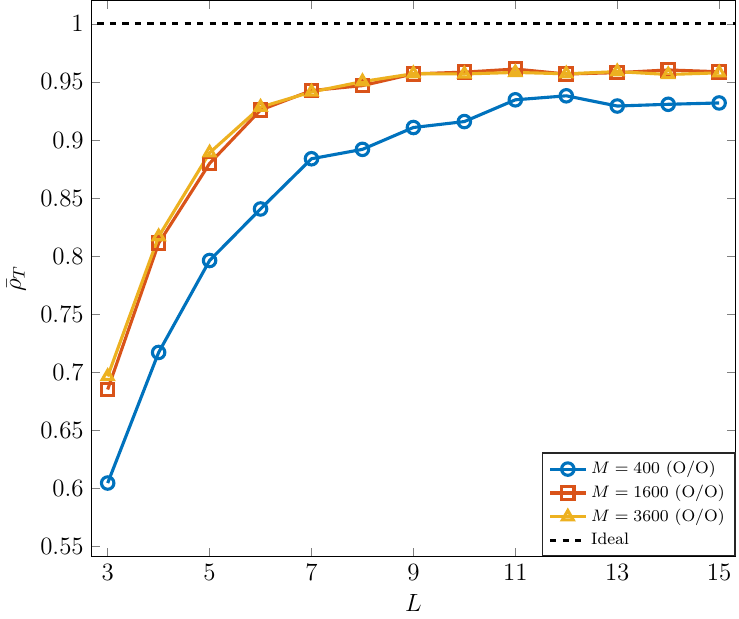}
        % \textbf{\hspace{10mm}(a)}
        \par\vspace{2pt}\hspace{10mm}\footnotesize\textbf{(a)}\par
    \end{minipage}%
    \hfill
    \begin{minipage}[t]{0.42\textwidth}
        \centering
        \includegraphics[width=\linewidth]{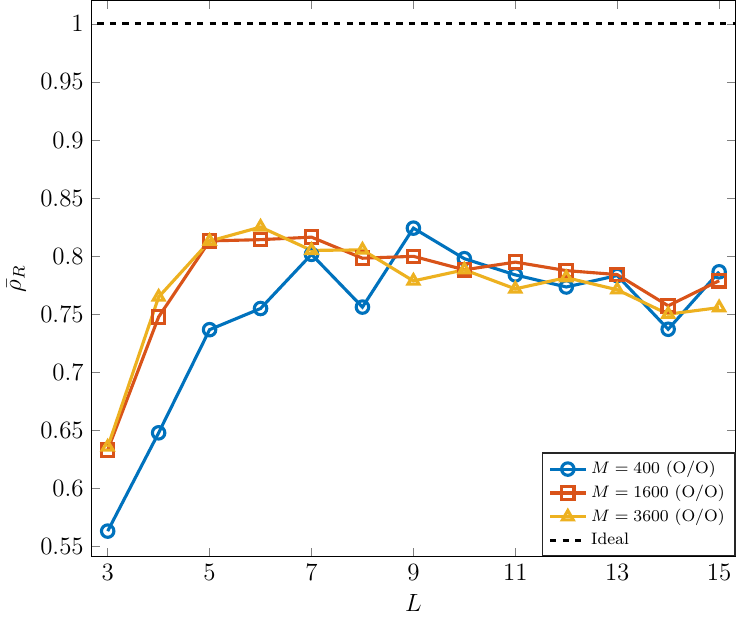}
        % \textbf{\hspace{10mm}(b)}
         \par\vspace{2pt}\hspace{10mm}\footnotesize\textbf{(b)}\par
    \end{minipage}
  \caption{Mean \ac{OAM} mode correlation coefficients versus the number of layers $L$ for the O/O configuration: (a)~transmit-side average correlation $\bar{\rho}_{\mathrm{T}}$ and (b)~receive-side average correlation $\bar{\rho}_{\mathrm{R}}$, reported for $M\in\{400,1600,3600\}$.}
    \label{fig:corr_vs_L}
\end{figure*}
%-----------------------------------------------
The proposed \ac{SIM}-aided \ac{HMIMO} system operates at carrier frequency $f_c = 28~\text{GHz}$ with a bandwidth $\Delta f = 120~\text{MHz}$,which is consistent with 5G NR \ac{mmWave} deployments~\cite{3gpp38104}. The transmit power is $\Pt = 30~\text{dBm}$, and the noise power spectral density is $\sigma_{\mathrm{w}}^2 = -174~\text{dBm/Hz}$.

Both the \ac{TX} and \ac{RX} antennas employ \acp{ULA} with $S=11$ elements, spaced by $\lambda/2$ and placed at a distance $5\lambda$ from the corresponding \ac{SIM}. The \ac{TX}- and \ac{RX}-\ac{SIM} use identical square apertures with side $L_x = L_y$, discretized into square meta-atoms of area $A_t = \lambda^2/4$. The number of layers is symmetric ($L = K$), with $L \in [3,\,15]$, and aperture size varies over $L_x \in \{10\lambda,\, 20\lambda,\, 30\lambda\}$, corresponding to $M \in \{400,\, 1600,\, 3600\}$ meta-atoms per layer.

Following the fixed-thickness convention in the \ac{SIM} literature~\cite{an2023stacked_holographic, liu2025stacked_Clerx}, the total thickness is fixed at $T_{\mathrm{SIM}} = 5\lambda$ for all configurations, and intra-layer spacing is set as $d_{\text{layer}} = T_{\mathrm{SIM}} / L$. This preserves the \ac{SIM} physical footprint while allowing finer analog processing resolution as $L$ increases. Consequently, the intra-layer spacing ranges from $d_{\text{layer}} \approx 1.67\lambda$ at $L = 3$ to $d_{\text{layer}} \approx 0.33\lambda$ at $L = 15$.\footnote{The condition $L \leq 15$ ensures that $d_{\text{layer}} \geq \lambda/4$, so inter-layer coupling remains negligible and the phase-only transmission model remains valid~\cite{an2023stacked_holographic}.}

Two representative deployment scenarios are considered:

\begin{itemize}
  \item \emph{\ac{SR} link}: the link distance $d_{\text{link}} = d_{\text{SR}}$
    is defined as the maximum distance that still guarantees at least $S$ spatial \ac{DOF} ($N_{\mathrm{DOF}} \geq S$, with $N_{\mathrm{DOF}}$ denoting the number of \ac{DOF} associated with the wireless channel). This ensures full near-field spatial multiplexing of all $S$ \ac{OAM} modes, thus representing a spatially rich regime in which inter-modal interference is the dominant performance bottleneck.

\item \emph{\ac{LR} link}: the link distance is set near the midpoint of the radiative near-field region, where $N_{\mathrm{DOF}} \leq S$\footnote{To ensure physical consistency across aperture sizes, a minimum \ac{DOF} constraint is enforced: if the midpoint yields $N_{\mathrm{DOF}} < 1$, the distance is reduced to the maximum value satisfying $N_{\mathrm{DOF}} \geq 1$.}. In this regime, the channel rank is insufficient to support all $S$ orthogonal spatial streams, and the multiplexing gain is progressively lost with increasing distance.

\end{itemize}
Specifically, the results presented in Secs.~\ref{sec:Q1} and~\ref{sec:Q3} are obtained for the \ac{SR} case, whereas the \ac{LR} regime is examined in Sec.~\ref{sec:distancesweep}.

\subsection{Performance Metrics}

The system performance is evaluated using both per-mode and system-level metrics, jointly capturing the \ac{OAM} mode-synthesis accuracy and end-to-end link efficiency.

At the transmit and receive sides, the fidelity of each synthesized mode relative to its ideal \ac{OAM}-based basis function is quantified through the mode correlation coefficients $\rho_{\mathrm{T},s}$ and $\rho_{\mathrm{R},s}$, defined as
\begin{align}
\rho_{\mathrm{T},s} &= \frac{\left|\mathbf{f}_s^{\mathrm{H}}\, \mathbf{a}_{\mathrm{T},s}\right|}{\left\|\mathbf{f}_s\right\|\left\|\mathbf{a}_{\mathrm{T},s}\right\|}\,, \; \;s = 1,2,  \dots, S,\\
\rho_{\mathrm{R},s} &= \frac{\left|\mathbf{a}_{\mathrm{R},s}\, \mathbf{e}_s^{\mathrm{H}}\right|}{\left\|\mathbf{a}_{\mathrm{R},s}\right\|\left\|\mathbf{e}_s\right\|}\,, \;\; s = 1, 2, \dots, S,
\end{align}
where $\mathbf{a}_{\mathrm{T},s}$ and $\mathbf{a}_{\mathrm{R},s}$ denote synthesized \ac{OAM} mode vectors generated by the optimized \ac{TX}- and \ac{RX}-\ac{SIM} configurations $\boldP^{\star}$ and $\boldQ^{\star}$, respectively. Vectors $\mathbf{f}_s$ and $\mathbf{e}_s$ are the target \ac{OAM}-based precoding and combining vectors obtained by discretizing ideal \ac{OAM} bases (Sec.~\ref{Sec:OAM_HMIMO_bases}). These coefficients lie in $[0,1]$, with a value of one indicating perfect matching between synthesized and ideal modal patterns. The average transmit- and receive-side correlations are
\begin{equation}
\bar{\rho}_{\mathrm{T}} = \frac{1}{S} \sum_{s=1}^{S} \rho_{\mathrm{T},s}\, ,\qquad \text{and} 
\qquad
\bar{\rho}_{\mathrm{R}} = \frac{1}{S} \sum_{s=1}^{S} \rho_{\mathrm{R},s} \,,
\end{equation}
which provide compact indicators of overall \ac{OAM} synthesis accuracy.

For each \ac{OAM}-multiplexed stream, the corresponding \ac{SINR} is also evaluated, accounting for the trade-off between the desired signal power, inter-modal interference, and noise at the \ac{RX}-\ac{ULA}. Specifically, the \ac{SINR} of the $s$th mode is given by
\begin{equation}\label{eq:SINRs}
\mathrm{SINR}_s =
\frac{
    \left| \left[ \HE \right]_{s,s} \right|^2 \, \mathbb{E}\!\left\{ |x_s|^2 \right\}
}{
    \displaystyle \sum_{k \neq s} \left| \left[ \HE \right]_{s,k} \right|^2 \, \mathbb{E}\!\left\{ |x_k|^2 \right\} + \Sigma_{\tilde{\mathrm{w}}}
}\,,
\quad s = 1, 2, \dots, S \,.
\end{equation}

%------ FIGURE: corr_per_mode (double column) ------
\begin{figure*}[t!]
    \centering
    \begin{minipage}[t]{0.42\textwidth}
        \centering
        \includegraphics[width=\linewidth]{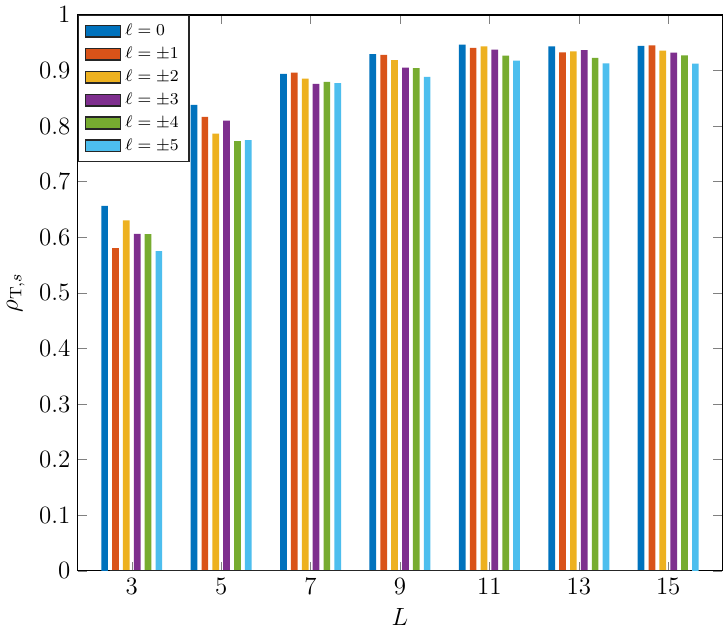}
      \par\vspace{2pt}\hspace{10mm}\footnotesize\textbf{(a)}\par
    \end{minipage}%
    \hfill
    \begin{minipage}[t]{0.42\textwidth}
        \centering
        \includegraphics[width=\linewidth]{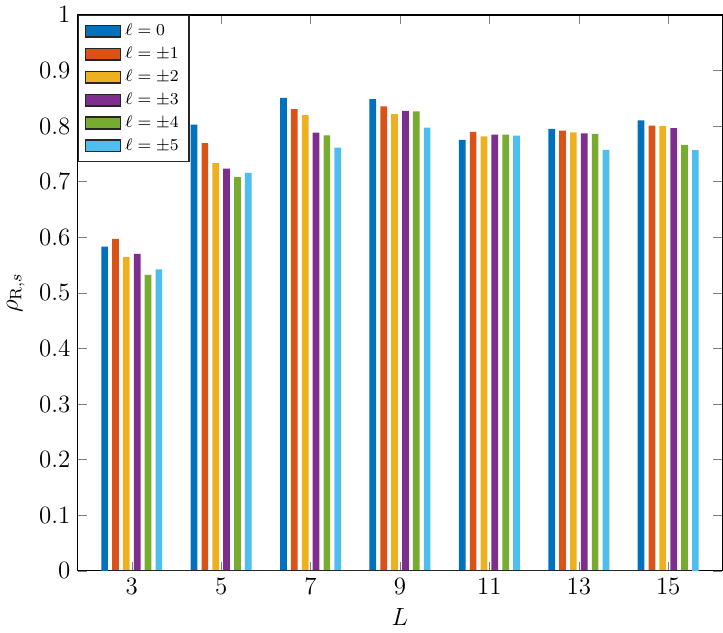}
         \par\vspace{2pt}\hspace{10mm}\footnotesize\textbf{(b)}\par
    \end{minipage}
\caption{Per-mode correlation coefficients for the O/O configuration with $M=400$ and odd layer sampling $L\in\{3,5,\dots,15\}$: (a)~transmit-side correlation $\rho_{\mathrm{T},s}$ and (b)~receive-side correlation $\rho_{\mathrm{R},s}$.}
    \label{fig:corr_per_mode}
\end{figure*}
%---------------------------------------------------

\begin{figure*}[t!]
\centering

% ---- GLOBAL SPACING CONTROL ----
\setlength{\tabcolsep}{0pt}
\setlength{\fboxsep}{0pt}

% ---------- HEADER ----------
\hspace{0.095\textwidth}
\begin{minipage}[c]{0.21\textwidth}\centering $\angle(\mathbf{F})$\end{minipage}
\hspace{0.01\textwidth}
\begin{minipage}[c]{0.21\textwidth}\centering $\angle(\boldA_{\mathrm{T}}),\, L=3$\end{minipage}
\hspace{0.01\textwidth}
\begin{minipage}[c]{0.21\textwidth}\centering $\angle(\boldA_{\mathrm{T}}),\, L=9$\end{minipage}

\vspace{2pt}

% ---------- ROW 1 ----------
\begin{minipage}[c][0.21\textwidth][c]{0.095\textwidth}
\centering $\ell_n=+3$
\end{minipage}
\begin{minipage}[c]{0.21\textwidth}
\includegraphics[width=\linewidth]{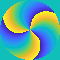}
\end{minipage}
\hspace{0.01\textwidth}
\begin{minipage}[c]{0.21\textwidth}
\includegraphics[width=\linewidth]{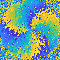}
\end{minipage}
\hspace{0.01\textwidth}
\begin{minipage}[c]{0.21\textwidth}
\includegraphics[width=\linewidth]{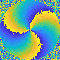}
\end{minipage}

\vspace{10pt}

% ---------- ROW 2 ----------
\begin{minipage}[c][0.21\textwidth][c]{0.095\textwidth}
\centering $\ell_n=+5$
\end{minipage}
\begin{minipage}[c]{0.21\textwidth}
\includegraphics[width=\linewidth]{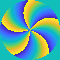}
\end{minipage}
\hspace{0.01\textwidth}
\begin{minipage}[c]{0.21\textwidth}
\includegraphics[width=\linewidth]{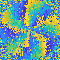}
\end{minipage}
\hspace{0.01\textwidth}
\begin{minipage}[c]{0.21\textwidth}
\includegraphics[width=\linewidth]{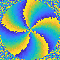}
\end{minipage}

\vspace{2pt}

\caption{Phase-profile matrices for the $\ell=+3$ and $\ell=+5$ \ac{OAM} modes. Columns show the ideal phase distribution $\angle(\mathbf{F})$ (left) and the synthesized transmit-side phase profile $\angle(\boldA_{\mathrm{T}})$ for $L=3$ (center) and $L=9$ (right).}
\label{fig:phase_matrix}

\vspace{-6pt} 
\end{figure*}

From a system-level perspective, the achievable link performance is characterized by the achievable sum-rate, 

representing the aggregate throughput across all $S$ modes, which is computed as
\begin{equation}\label{eq:Gamma_capacity}
\Gamma = \sum_{s=1}^{S} \log_{2}\,\bigl(1 + \mathrm{SINR}_s\bigr)\,.
\end{equation}

In addition, the following configurations are considered:

\begin{itemize}
    \item \textit{TX- and RX-optimized} (O/O): Both the transmitting and receiving \acp{SIM} are optimized according to Algorithm~\ref{alg:SIMopt}. This is the most realistic implementation.
  
    \item \textit{TX-only optimized} (O/I): The transmitting \ac{SIM} is optimized via Algorithm~\ref{alg:SIMopt}, whereas an ideal \ac{OAM} combiner is assumed at reception, i.e., $\Hr \boldQ = \mathbf{E}$. This isolates the impact of \ac{TX}-\ac{SIM} optimization on \ac{OAM} mode generation.
  
    \item \textit{RX-only optimized} (I/O): An ideal \ac{OAM} precoder is assumed, that is, the \ac{TX} \ac{ULA}-\ac{SIM} cascade $\boldP \Ht$ is replaced by target matrix $\mathbf{F}$, while the receiving \ac{SIM} is optimized by the proposed correlation-driven algorithm.
    
    \item \textit{Ideal}: Both \ac{OAM} precoding and combining applied via ideal matrices $\mathbf{F}$ and $\mathbf{E}$, thereby providing an upper-bound benchmark.
\end{itemize}

%--------------------------------------------------------------------------
\subsection{Impact of SIM Architecture on OAM Mode Fidelity}\label{sec:Q1}
%--------------------------------------------------------------------------

%------ FIGURE: sinr_per_mode (double column) ------
\begin{figure*}[t!]
    \centering
    \begin{minipage}[t]{0.31\textwidth}
        \centering
        \includegraphics[width=\linewidth]{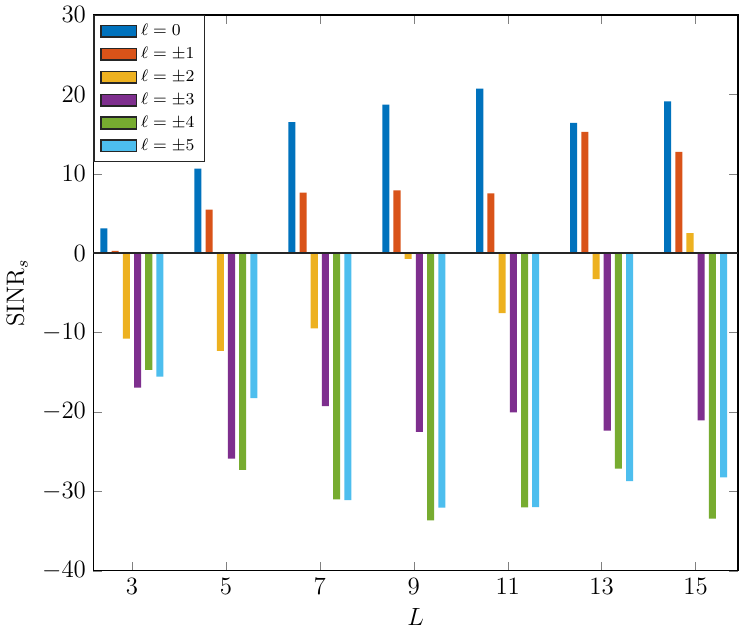}
        \par\vspace{2pt}\hspace{8mm}\footnotesize\textbf{(a)}\par
    \end{minipage}%
    \hfill
    \begin{minipage}[t]{0.31\textwidth}
        \centering
        \includegraphics[width=\linewidth]{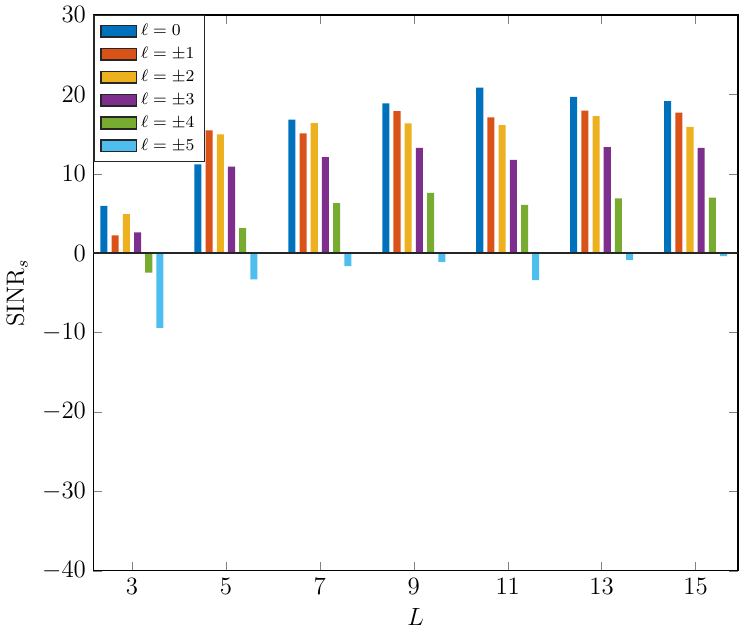}
        \par\vspace{2pt}\hspace{8mm}\footnotesize\textbf{(b)}\par
    \end{minipage}%
    \hfill
    \begin{minipage}[t]{0.31\textwidth}
        \centering
        \includegraphics[width=\linewidth]{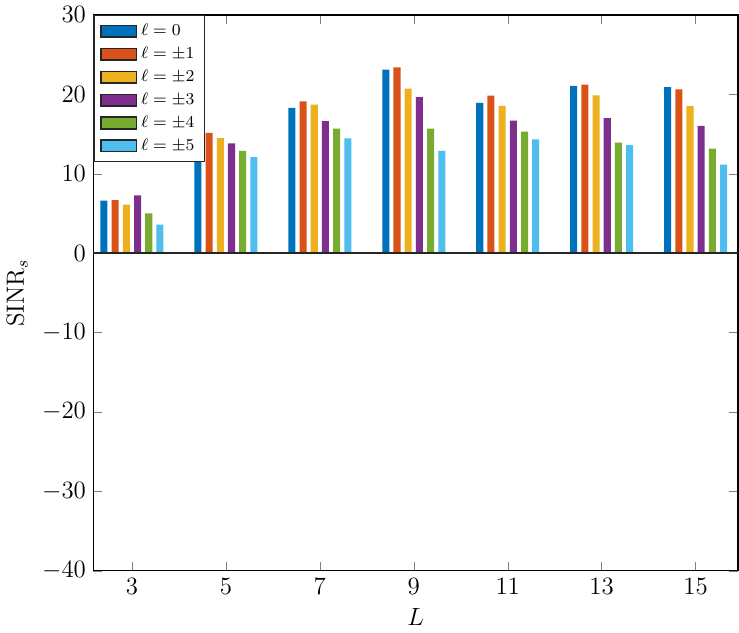}
        \par\vspace{2pt}\hspace{8mm}\footnotesize\textbf{(c)}\par
    \end{minipage}
\caption{Per-mode \ac{SINR} versus the number of layers $L$ for the O/O configuration and $\ell = 0,\pm1,\dots,\pm5$: (a)~$M=400$, (b)~$M=1600$, and (c)~$M=3600$.}
    \label{fig:sinr_per_mode}
\end{figure*}
%---------------------------------------------------

In this subsection, we quantify the impact of the number of layers $L$ and meta-atom density $M$ on the \ac{OAM} mode synthesis accuracy under the \ac{SR} regime, thereby isolating the architectural effects from \ac{SNR}-limited impairments.

Fig.~\ref{fig:corr_vs_L}(a) and Fig.~\ref{fig:corr_vs_L}(b) show the average transmit- and receive-side correlations, i.e., $\bar{\rho}_{\mathrm{T}}$ and $\bar{\rho}_{\mathrm{R}}$, versus $L$ for different $M$ values. Two key effects emerge. First, increasing $L$ consistently improves the performance, confirming the benefit of deeper \ac{SIM} stacks in enhancing the \ac{EM} wave transformation fidelity. However, the gain rapidly saturates, indicating diminishing returns once the stack capacity sufficiently resolves the target phase transformation. Indeed, it reaches $\bar{\rho}_{\mathrm{T}} \approx 0.95$ for $M=3600$ for $L \geq 7$. Second, increasing $M$ yields a clear and systematic improvement across all $L$. This is directly linked to finer spatial sampling of the \ac{OAM} mode wavefront: a higher meta-atom density enables more accurate discretization of the characteristic helical phase profile, improving the mode reconstruction fidelity. Conversely, low-$M$ regimes are fundamentally under-sampled, limiting performance regardless of the \ac{SIM} stack depth. The receive-side metric in Fig.~\ref{fig:corr_vs_L}(b) follows the same general trend, although it exhibits lower and more variable $\bar{\rho}_{\mathrm{R}}$ values. This is due to error accumulation: $\bar{\rho}_{\mathrm{R}}$ reflects both the imperfect transmit-side \ac{OAM} synthesis and the residual mismatch at the \ac{RX}-\ac{SIM}; therefore any degradation at the \ac{TX} side propagates into the receive-side metric.

Furthermore, Fig.~\ref{fig:corr_per_mode} illustrates the per-mode correlation coefficients as a function of $L$. In particular, Fig.~\ref{fig:corr_per_mode}(a) shows $\rho_{\mathrm{T},s}$, while Fig.~\ref{fig:corr_per_mode}(b) shows $\rho_{\mathrm{R},s}$, both evaluated for the O/O configuration across different topological charges and for $M = 400$. Overall, increasing $L$ consistently enhances the \ac{OAM} mode generation fidelity, independent of the topological charge. 

In fact, low-order modes tend to exhibit slightly higher fidelity for small $L$, although the difference across topological charges is not pronounced. As $\ell$ increases, deeper \ac{SIM} architectures become increasingly beneficial, since higher-order \ac{OAM} phase profiles impose more complex wavefront transformations. Overall, increasing $L$ consistently enhances the \ac{OAM} mode generation fidelity across all topological charges, with the most evident improvements visible in the synthesized phase profiles. In this regard, Fig.~\ref{fig:phase_matrix} compares ideal and synthesized phase profiles for $\ell = +3$ and $\ell = +5$ modes at $L=3$ and $L=9$. At $L=3$, the reconstructed phase distributions exhibit significant noise and spatial fragmentation. In contrast, increasing the stack depth to $L=9$ yields a markedly improved reconstruction, where the characteristic helical phase structure is clearly generated, with sharper phase transitions and enhanced rotational symmetry.

%--------------------------------------------------------------------------
\subsection{SINR Performance and Spectral Efficiency}\label{sec:Q3}
%--------------------------------------------------------------------------

Fig.~\ref{fig:sinr_per_mode} illustrates the per-mode \ac{SINR} as a function of $L$ for different aperture sizes $M$ and topological charges in the O/O configuration. Fig.~\ref{fig:sinr_per_mode}(a), (b), and (c) correspond to $M = 400$, $1600$, and $3600$, respectively.

For $M=400$, a pronounced inter-mode imbalance is observed: aside from the fundamental mode ($\ell=0$), most \ac{OAM} modes exhibit negative \ac{SINR} across all $L$ values, indicating a regime dominated by inter-mode interference. While increasing $L$ yields a uniform upward shift in performance, the overall imbalance persists, highlighting a fundamental limitation imposed by the reduced \ac{SIM} aperture that cannot be compensated for solely by increasing the stack depth. For $M=1600$, a more balanced behavior emerges, with the majority of \ac{OAM} modes achieving positive \ac{SINR}. In this regime, increasing $L$ provides consistent performance gains across modes, indicating improved controllability of the spatial \ac{DOF}.

A similar trend is observed for $M=3600$, where overall performance is significantly enhanced. An extra gain in the order of $10$-$15\,\mathrm{dB}$ is observed for higher-order modes compared to the smaller aperture cases. Notably, the benefits of increasing $M$ are most pronounced for higher-order modes, confirming that the aperture size plays a critical role in supporting the synthesis of spatially complex OAM profiles and mitigating inter-mode interference.

\begin{figure}[t!]
    \centering
    \includegraphics[width=0.42\textwidth]{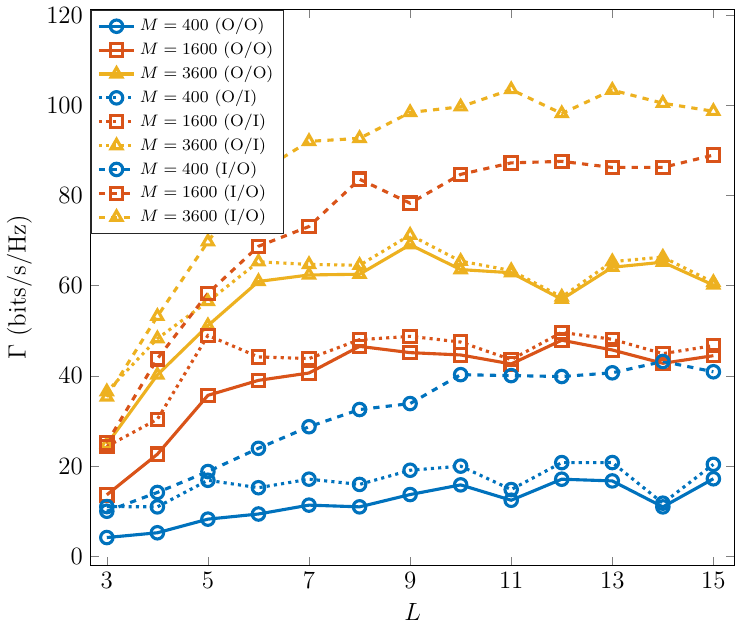}
    \caption{Achievable sum-rate $\Gamma$ versus the number of layers $L$ for different aperture sizes ($M=400,1600,3600$) and optimization configurations (O/O, O/I, I/O).}
    \label{fig:capacity_vs_L}
\end{figure}

In addition, Fig.~\ref{fig:capacity_vs_L} shows the achievable sum-rate $\Gamma$ versus the number of layers $L$ for $M \in \{400, 1600, 3600\}$ under the three optimization configurations (O/O, O/I, I/O). The O/O case exhibits a pronounced performance gap with respect to both partially idealized settings, particularly the I/O case, whose substantial gain underscores the critical role of accurate transmit-side \ac{OAM} synthesis. Imperfections in the radiated field distribution fundamentally constrain multiplexing performance, regardless of the receiver-side processing quality. As a function of $L$, $\Gamma$ grows steeply for small-to-moderate depths and saturates beyond $L \approx 8$, indicating that additional layers yield diminishing returns once a sufficient phase-shaping capability is available within the \ac{SIM} stack. A strong aperture-scaling trend is consistently observed across all configurations: a larger $M$ monotonically increases capacity, with the most pronounced effect in the I/O case, where $\Gamma$ reaches approximately $100$~bits/s/Hz for $M = 3600$, confirming that large apertures are essential to fully exploit the near-field channel \acp{DOF} offered by \ac{HMIMO} systems.

%--------------------------------------------------------------------------
\subsection{Impact of Link Distance}\label{sec:distancesweep}
\begin{figure}[t!]
    \centering
    \includegraphics[width=0.42\textwidth]{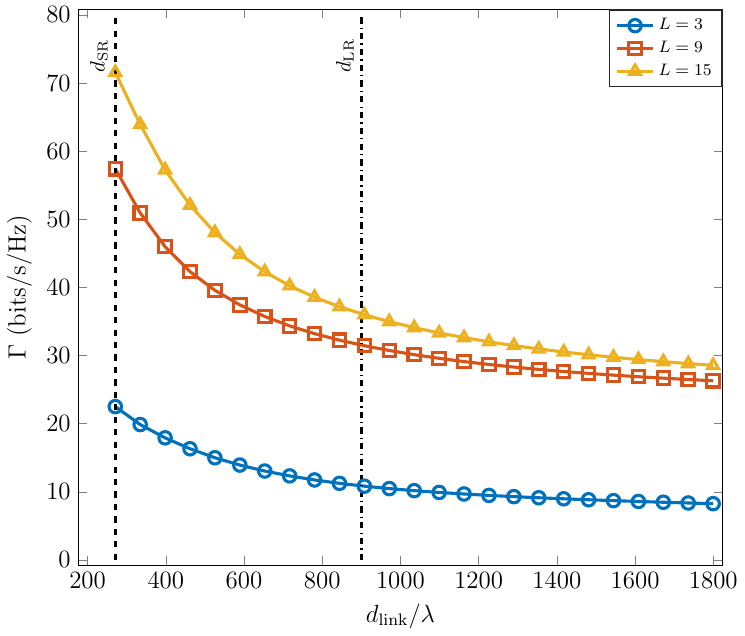}
\caption{Achievable sum-rate $\Gamma$ versus normalized link distance $d_{\mathrm{link}}/\lambda$ for $L\in\{3,9,15\}$ and $M=3600$. The reference distances $d_{\mathrm{SR}}$ and $d_{\mathrm{LR}}$ mark the boundaries of the \ac{SR} and \ac{LR} propagation regimes.}
\label{fig:capacity_vs_distance}
\end{figure}

%--------------------------------------------------------------------------
Fig.~\ref{fig:capacity_vs_distance} illustrates the sum-rate $\Gamma$ versus the normalized \ac{HMIMO} link distance $d_{\text{link}}/\lambda$ for $L =\{3,9, 15\}$. In this case, both the TX and RX SIMs are optimized to support \ac{OAM}-based spatial multiplexing at a distance $d_{\text{link}} = d_{\text{LR}}$. 

Three distinct operating regimes are identified in the figure. For $d_{\text{link}} \leq d_{\text{SR}}$, the near-field \ac{HMIMO} channel supports a large number of orthogonal communication modes, thereby enabling full multiplexing with well-coupled \ac{OAM} modes; deeper \ac{SIM} architectures yield substantially higher capacity owing to superior inter-mode interference suppression. 
In the transition region ($d_{\text{SR}} \leq d_{\text{link}} \leq d_{\text{LR}}$), $\Gamma$ degrades monotonically as the channel rank decreases with increasing distance, with the steepest drop observed for large $L$ due to the higher multiplexing gain being progressively lost. Beyond $d_{\text{LR}}$, the \ac{LOS} channel rank approaches unity, propagation-induced constraints dominate, and all curves converge toward a \ac{DOF}-limited floor. 

It is worth noting that in all the considered configurations, the \ac{SIM} phases are optimized at $d_{\text{link}} = d_{\text{LR}}$ and remained fixed across the entire distance range. The graceful degradation observed in Fig.~\ref{fig:capacity_vs_distance} thus confirms that the proposed correlation-driven design is robust to link-distance mismatch: once optimized for a target operating point, the \ac{SIM} configuration does not require distance-specific re-optimization. This robustness stems from the inter-modal crosstalk suppression capability of deeper stacks, which remains effective even as the channel rank decreases with distance.

%
%------ FIGURE: capacity_vs_L (single column) ------
%---------------------------------------------------

%--------------------------------------------------------------------------
\subsection{Algorithm Convergence}\label{sec:convergence}
\begin{figure}[t!]
    \centering
    \hspace{-5mm}
    \includegraphics[width=0.44\textwidth]{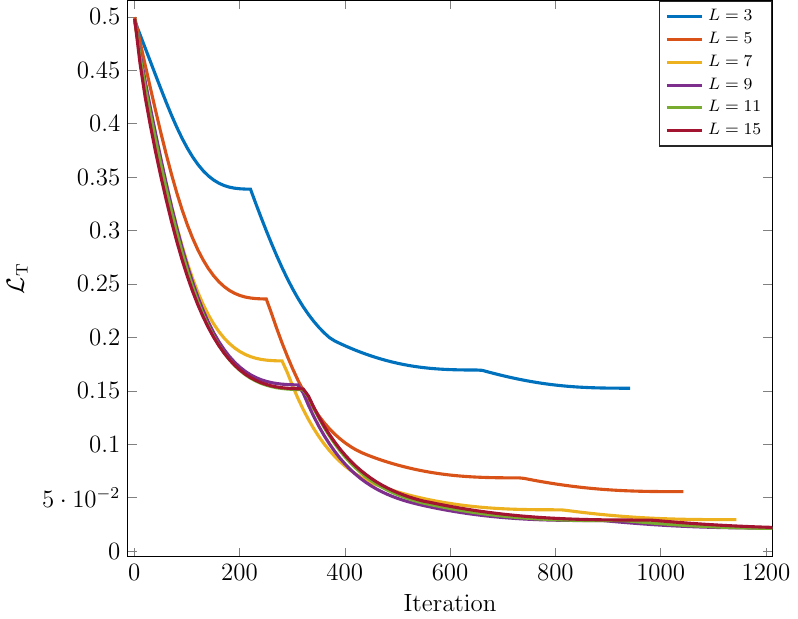}
\caption{Convergence of the transmit-side optimization loss $\mathcal{L}_{\mathrm{T}}$ as a function of iteration index for $M=3600$ and different $L$ values.}
    \label{fig:loss_convergence}
\end{figure}

%--------------------------------------------------------------------------
Fig.~\ref{fig:loss_convergence} illustrates the convergence profile of the transmit-side loss $\mathcal{L}_{\mathrm{T}}$ versus the iteration count for varying $L$. All the configurations converge monotonically from random initialization and stabilize within approximately $1200$ iterations. A larger $L$ consistently achieves a lower final loss and faster convergence, indicating a more favorable optimization landscape attributable to the increased \ac{DOF} and enhanced \ac{EM} wavefront shaping flexibility of the deeper stacks. These results confirm that the proposed multi-start Adam strategy yields a reliable convergence across all considered configurations.

%------ FIGURE: capacity_vs_distance (single column) ------
%----------------------------------------------------------
%------ FIGURE: loss_convergence (single column) ------
%------------------------------------------------------

%--------------------------------------------------------------------------
\section{Conclusion}\label{sec:Conclusion}
%-------------------------------------------------------------------------
This study proposes a correlation-driven optimization framework for \ac{OAM}-enabled \ac{HMIMO} communications via \acp{SIM}-based transceivers, enabling robust wave-domain synthesis and demultiplexing of orthogonal \ac{OAM} modes in compact discrete architectures.

Numerical results reveal a clear decoupling between the \ac{SIM} aperture, which limits the maximum supportable mode order via spatial sampling, and the layer depth, which mainly governs inter-modal crosstalk suppression with rapidly diminishing returns beyond moderate stack sizes. Specifically, deploying denser apertures significantly sharpens the per-mode correlation, whereas expanding the layer count primarily enhances the system robustness against phase discretization constraints.

Transmit/receive asymmetries significantly affect the performance, with radiated-field imperfections causing substantially stronger throughput loss than non-ideal reception, highlighting the critical role of transmit-side mode synthesis. Overall, \ac{SIM}-based \ac{HMIMO} transceivers enable efficient near-field spatial multiplexing and exhibit graceful distance-dependent degradation without requiring distance-specific \ac{SIM} re-optimization.
These findings provide practical design guidelines for future \ac{SIM}-based \ac{HMIMO} transceivers operating in the radiative near field.

\section*{Acknowledgments}

G. Torcolacci and D. Dardari are with the Department of Electrical, Electronic, and Information Engineering “Guglielmo Marconi”, University of Bologna, Italy, and CNIT-WiLab, Bologna, Italy (e-mail: \{g.torcolacci, davide.dardari\}@unibo.it). \\
G. Torcolacci was funded by an NRRP Ph.D. grant.\\
\textit{This work has been submitted to IEEE for possible publication. Copyright may be transferred without notice.}

\bibliographystyle{IEEEtran}
\bibliography{references}

@techreport{3gpp38104,
  author      = {{3GPP}},
  title       = {{NR; Base Station (BS) radio transmission and reception}},
  institution = {{3rd Generation Partnership Project}},
  type        = {{Technical Specification (TS)}},
  number      = {38.104},
  version     = {18.5.0},
  year        = {2024},
  note        = {Release~18}
}

@article{AnXuetAl2:J23,
  title={Stacked intelligent metasurface-aided {MIMO} transceiver design},
  author={An, Jiancheng and Yuen, Chau and Xu, Chao and Li, Hongbin and Ng, Derrick Wing Kwan and Di Renzo, Marco and Debbah, M{\'e}rouane and Hanzo, Lajos},
  journal={IEEE Wireless Commun.},
  volume={31},
  number={4},
  pages={123--131},
  year={2024},
month={Apr.},
  publisher={IEEE}
}

@article{khan2026holistic,
  title={A Holistic Review of Holographic {MIMO} for {6G}: Hardware Design and Implementation Perspectives},
  author={Khan, Daud and Ullah, Arif and Din, Iftikhar Ud and Ouameur, Messaoud Ahmed and Bagaa, Miloud and Massicotte, Daniel and Razmpoosh, Bahram},
journal={IEEE Open J. Commun. Soc.},
  year={2026},
  month={Apr.},
  publisher={IEEE}
}

@book{rudin1987real,
  author    = {Rudin, Walter},
  title     = {Real and Complex Analysis},
  edition   = {3},
  publisher = {McGraw-Hill},
  year      = {1987},
}

@inproceedings{
loshchilov2017sgdr,
title={{SGDR}: Stochastic Gradient Descent with Warm Restarts},
author={Ilya Loshchilov and Frank Hutter},
booktitle={Proc. Int. Conf. Learn. Represent. (ICLR)},
year={2017},
month={Jun.}
}

@ARTICLE{an2023multiuser,
  author={An, Jiancheng and Di Renzo, Marco and Debbah, Mérouane and Vincent Poor, H. and Yuen, Chau},
  journal={IEEE Trans. Wireless Commun.}, 
  title={Stacked Intelligent Metasurfaces for Multiuser Downlink Beamforming in the Wave Domain}, 
  year={2025},
  volume={24},
  month={Mar.},
  number={7},
  pages={5525-5538}}

@article{LiuetAl:J22,
  title={A programmable diffractive deep neural network based on a digital-coding metasurface array},
  author={Liu, Che and Ma, Qian and Luo, Zhang Jie and Hong, Qiao Ru and Xiao, Qiang and Zhang, Hao Chi and Miao, Long and Yu, Wen Ming and Cheng, Qiang and Li, Lianlin and others},
  journal={Nature Electronics},
  volume={5},
  number={2},
  pages={113--122},
month={Feb.},
  year={2022},
  publisher={Nature Publishing Group UK London}
}

@book{Bal:B15,
  title={Antenna Theory: Analysis and Design},
  author={Balanis, C.A.},
  isbn={9781119178989},
  lccn={2016050162},
  year={2015},
  publisher={Wiley}
}

@ARTICLE{TorDecDar:J23,
  author={Torcolacci, Giulia and Decarli, Nicolò and Dardari, Davide},
  journal={IEEE Open J. Commun. Soc.}, 
  title={Holographic {MIMO} Communications Exploiting the Orbital Angular Momentum}, 
  year={2023},
  volume={4},
  number={},
  pages={1452-1469},
month={Jul.},
  doi={10.1109/OJCOMS.2023.3293197}}

@inproceedings{wang2024multi,
  title={Multi-user {ISAC} through stacked intelligent metasurfaces: New algorithms and experiments},
  author={Wang, Ziqing and Liu, Hongzheng and Zhang, Jianan and Xiong, Rujing and Wan, Kai and Qian, Xuewen and Di Renzo, Marco and Qiu, Robert Caiming},
  booktitle={Proc. IEEE Global Commun. Conf. (GLOBECOM)},
  pages={4442--4447},
  month={Dec.},
  year={2024}
}

@inproceedings{vanwynsberghe2023walsh,
  title={Walsh Meets {OAM} in Holographic {MIMO}},
  author={Vanwynsberghe, Charles and He, Jiguang and Huang, Chongwen and Debbah, Merouane},
  booktitle={Int. Conf. Electromagn. Adv. Appl. (ICEAA)},
  pages={593--596},
  year={2023},
month={Oct.},
  organization={IEEE}
}

@article{gong2023holographic,
  title={Holographic {MIMO} communications: Theoretical foundations, enabling technologies, and future directions},
  author={Gong, Tierui and Gavriilidis, Panagiotis and Ji, Ran and Huang, Chongwen and Alexandropoulos, George C and Wei, Li and Zhang, Zhaoyang and Debbah, M{\'e}rouane and Poor, H Vincent and Yuen, Chau},
  journal={IEEE Commun. Surv. Tutorials.},
month={Aug.},
  year={2023},
  publisher={IEEE}
}

@article{dardari2020communicating,
  title={Communicating with large intelligent surfaces: Fundamental limits and models},
  author={Dardari, Davide},
  journal={IEEE J. Sel. Areas Commun.},
  volume={38},
  number={11},
  pages={2526--2537},
month={Nov.},
  year={2020},
  publisher={IEEE}
}

@article{torcolacci2024holographic,
  title={Holographic imaging with {XL-MIMO} and {RIS}: Illumination and reflection design},
  author={Torcolacci, Giulia and Guerra, Anna and Zhang, Haiyang and Guidi, Francesco and Yang, Qianyu and Eldar, Yonina C and Dardari, Davide},
  journal={IEEE J. Sel. Top. Signal Process.},
  year={2024},
    volume={18},
  number={4},
  pages={587-602},
  month={Jun.},
  publisher={IEEE}
}

@article{decarli2021communication,
  title={Communication modes with large intelligent surfaces in the near field},
  author={Decarli, Nicol{\`o} and Dardari, Davide},
  journal={IEEE Access},
  volume={9},
month={Dec.},
  pages={165648--165666},
  year={2021},
  publisher={IEEE}
}

@article{chen2019orbital,
  title={Orbital angular momentum waves: Generation, detection, and emerging applications},
  author={Chen, Rui and Zhou, Hong and Moretti, Marco and Wang, Xiaodong and Li, Jiandong},
  journal={IEEE Commun. Surv. Tutorials.},
  volume={22},
  number={2},
month={Nov.},
  pages={840--868},
  year={2019},
  publisher={IEEE}
}

@article{xie2015performance,
  title={Performance metrics and design considerations for a free-space optical orbital-angular-momentum--multiplexed communication link},
  author={Xie, Guodong and Li, Long and Ren, Yongxiong and Huang, Hao and Yan, Yan and Ahmed, Nisar and Zhao, Zhe and Lavery, Martin PJ and Ashrafi, Nima and Ashrafi, Solyman and others},
  journal={Optica},
  volume={2},
  number={4},
  pages={357--365},
month={Apr.},
  year={2015},
  publisher={Optica Publishing Group}
}

@inproceedings{AnXuetAl3:J23,
  title={Stacked intelligent metasurface performs a {2D} {DFT} in the wave domain for {DOA} estimation},
  author={An, Jiancheng and Yuen, Chau and Di Renzo, Marco and Debbah, Merouane and Poor, H Vincent and Hanzo, Lajos},
booktitle={Proc. IEEE Int. Conf. Commun. (ICC)},
  pages={3445--3451},
  month={Jun.},
  year={2024}
}

@ARTICLE{niu2025meta,
  author={Niu, Hong and An, Jiancheng and Wu, Tuo and Chen, Jiangong and Zhao, Yufei and Liang Guan, Yong and Di Renzo, Marco and Debbah, Mérouane and Karagiannidis, George K. and Vincent Poor, H. and Yuen, Chau},
  journal={IEEE Trans. Wireless Commun.}, 
  title={Introducing Meta-Fiber Into Stacked Intelligent Metasurfaces for {MIMO} Communications: A Low-Complexity Design With Only Two Layers}, 
  year={2026},
  volume={25},
  number={},
  month={Aug.},
  pages={3016-3032}}

@article{liu2025stacked_Clerx,
  title={Stacked intelligent metasurfaces for wireless communications: Applications and challenges},
  author={Liu, Hao and An, Jiancheng and Jia, Xing and Gan, Lu and Karagiannidis, George K and Clerckx, Bruno and Bennis, Mehdi and Debbah, M{\'e}rouane and Cui, Tie Jun},
  journal={IEEE Wireless Commun.},
  volume={32},
  month={Aug.},
  number={4},
  pages={46--53},
  year={2025},
  publisher={IEEE}
}

@article{YaoetAl:J24,
  title={Channel Estimation for Stacked Intelligent Metasurface-Assisted Wireless Networks},
  author={Yao, Xianghao and An, Jiancheng and Gan, Lu and Di Renzo, Marco and Yuen, Chau},
  journal={IEEE Wirel. Commun.},
  year={2024},
   volume={13},
  number={5},
  pages={1349-1353},
month={Feb.},
  publisher={IEEE}
}

@article{an2023stacked_holographic,
  author={An, Jiancheng and Xu, Chao and Ng, Derrick Wing Kwan and Alexandropoulos, George C. and Huang, Chongwen and Yuen, Chau and Hanzo, Lajos},
  journal={IEEE J. Sel. Areas Commun.}, 
  title={Stacked Intelligent Metasurfaces for Efficient Holographic {MIMO} Communications in {6G}}, 
  year={2023},
  volume={41},
  number={8},
  pages={2380-2396},
  month={Jun.},
  doi={10.1109/JSAC.2023.3288236}
}

@article{darsena2025design_amplitude,
  author={Darsena, Donatella and Verde, Francesco and Iudice, Ivan and Galdi, Vincenzo},
  journal={IEEE Open Journal of the Communications Society},
  title={Design of Stacked Intelligent Metasurfaces With Reconfigurable Amplitude and Phase for Multiuser Downlink Beamforming},
  year={2025},
  volume={6},
  pages={1021-1036},
  month={Jan.},
  doi={10.1109/OJCOMS.2025.3527094}
}

@article{papazafeiropoulos2024achievable,
  author={Papazafeiropoulos, Anastasios K. and An, Jiancheng and Kourtessis, Pandelis and Ratnarajah, Tharmalingam and Chatzinotas, Symeon},
journal={IEEE Trans. Wireless Commun.},
  title={Achievable Rate Optimization for Stacked Intelligent Metasurface-Assisted Holographic {MIMO} Communications}, 
  year={2024},
  volume={23},
  number={10},
  pages={13173-13186},
  month={Oct.},
  doi={10.1109/TWC.2024.3382990}
}

@inproceedings{liu2024drl_multiuser,
  author={Liu, Hongbin and An, Jiancheng and Ng, Derrick Wing Kwan and Alexandropoulos, George C. and Gan, Lu},
  booktitle={Proc. IEEE Int. Conf. Commun. (ICC)}, 
  title={{DRL}-Based Orchestration of Multi-User {MISO} Systems with Stacked Intelligent Metasurfaces}, 
  year={2024},
  pages={1713-1718},
  month={June},
  doi={10.1109/ICC51166.2024.10622418}
}

@article{HassanetAl:J24,
  title={Efficient beamforming and radiation pattern control using stacked intelligent metasurfaces},
  author={Hassan, Naveed Ul and An, Jiancheng and Di Renzo, Marco and Debbah, M{\'e}rouane and Yuen, Chau},
  journal={IEEE Open J. Commun. Soc.},
month={Jan.},
  year={2024},
  publisher={IEEE}
}

@article{lee2024continuous,
  title={Continuous-Aperture {OAM} Communication Systems: An Electromagnetic Information Theory Perspective},
  author={Lee, Young-Seok and Kim, Young Dam and Jung, Bang Chul},
  journal={IEEE Wireless Commun. Lett.},
  year={2024},
    volume={14},
  number={3},
  pages={676-680},
  publisher={IEEE},
month={Mar.},
}

@inproceedings{dai2024optimal,
  title={Optimal Mode for Continuous Aperture {MIMO} Based Holographic Communications},
  author={Dai, Weijie and Zhang, Yize and Zhang, Liyan and Tang, Xinke and Song, Jian and Dong, Yuhan},
booktitle={Proc. IEEE Int. Conf. Commun. Workshops (ICC Workshops)},
  pages={1938--1943},
month={Jun.},
  year={2024},
  organization={IEEE}
}

@article{jin2025achieving,
  title={Achieving High-Capacity {OAM} Communication With Fluid-Antenna-Based Continuous-Aperture Arrays},
  author={Jin, Hongyun and Cheng, Wenchi and Wang, Jingqing and Du, Qinghe and Zhang, Wei},
  journal={IEEE J. Sel. Areas Commun.},
  year={2025},
month={Sept.}, 
volume={44},
  number={},
  pages={1449-1463},
  publisher={IEEE}
}

@ARTICLE{Xu:J17,
  author={Xu, Jie},
  journal={IEEE Trans. Antennas Propag.}, 
  title={Degrees of Freedom of {OAM}-Based Line-of-Sight Radio Systems}, 
  year={2017},
  volume={65},
  number={4},
  pages={1996-2008},
  month = {Apr.},
  publisher={IEEE},
  doi={10.1109/TAP.2017.2671430}}

@inproceedings{niu2024physical_security,
  author={Niu, Hongyan and An, Jiancheng and Zhang, Lei and Lei, Xianfu and Yuen, Chau},
booktitle={Proc. IEEE VTS Asia Pac. Wirel. Commun. Symp. (APWCS)},
  title={Enhancing Physical Layer Security for {SISO} Systems Using Stacked Intelligent Metasurfaces}, 
  year={2024},
  pages={223-228},
  month={Aug.}
}

@article{huang2024taskoriented,
  author={Huang, Gaojie and An, Jiancheng and Yang, Zhaolin and Gan, Lu and Bennis, Mehdi and Debbah, Merouane},
  journal={IEEE Wireless Commun. Lett.},
  title={Stacked Intelligent Metasurfaces for Task-Oriented Semantic Communications},
  year={2025},
  volume={14},
  number={2},
  pages={310-314},
  month={Feb.},
  doi={10.1109/LWC.2024.3431001}
}

@article{fan2024multiplexed_oam,
  author={Fan, Yuan and Yang, Yuhan and Wang, Jiaxin and Chen, Rui and Xie, Yuanmu and Liu, Hao and Tang, Jie and Wang, Zhonglei and Li, Junjie and Ding, Ruwen},
  journal={Nature Communications},
  title={Multiplexed Manipulation of Orbital Angular Momentum and Amplitude Modulation},
  year={2024},
  volume={15},
  pages={3150},
  month={Apr.},
  doi={10.1038/s41467-024-47433-4}
}

@article{Mil:C00,
  title={Communicating with waves between volumes: evaluating orthogonal spatial channels and limits on coupling strengths},
  author={Miller, David AB},
  journal=APP_OPT,
  volume={39},
  number={11},
  pages={1681--1699},
  year={2000},
  month={Apr.},
  publisher={Optical Society of America}
}

@article{dardari2024overview,
  title={Over-the-Air Electromagnetic Signal Processing: From fundamentals to emerging paradigms},
  author={Dardari, Davide and Torcolacci, Giulia and Pasolini, Gianni and Decarli, Nicol{\`o}},
  journal={IEEE Signal Process. Mag.},
  volume={43},
  number={1},
  month={Feb.},
  pages={6--28},
  year={2026},
  publisher={IEEE}
}

@STRING{Globecom = {Proc. IEEE Global Telecomm. Conf.}}

@STRING{ICC = {Proc. IEEE Int. Conf. on Commun.}}

@STRING{IEEE = {The Institute of Electrical and Electronics Engineers,}}

@STRING{JIN = {J. of the Institute of Navigation}}

@STRING{APP_OPT = {Appl. Opt.}}

\end{document}